\documentclass[aip,jcp,12pt,a4paper]{revtex4-1}
\pdfoutput=1
\usepackage{graphicx}
\usepackage{amsmath,amssymb}
\usepackage{subfigure}

\newcommand{\symbf}{\boldsymbol}

\begin{document}

\title{The Exact Wavefunction Factorization of a Vibronic Coupling System}

\author{Ying-Chih Chiang$^1$, Shachar Klaiman$^1$, Frank Otto$^1$,  Lorenz S. Cederbaum$^1$}

\affiliation{$^1$ Theoretische Chemie, Universit\"at Heidelberg,
    Im Neuenheimer Feld 229, D--69120 Heidelberg, Germany}

\date{\today}

\begin{abstract} 
We investigate the exact wavefunction as a single product of 
electronic and nuclear wavefunction for a model conical intersection system.
Exact factorized spiky potentials and nodeless nuclear wavefunctions are
found. The exact factorized potential preserves the symmetry breaking effect
when the coupling mode is present. Additionally the nodeless wavefunctions
are found to be closely related to the adiabatic nuclear eigenfunctions.
This phenomenon holds even for the regime where the non-adiabatic coupling
is relevant, and sheds light on the relation between the exact wavefunction 
factorization and the adiabatic approximation.
\end{abstract}

\maketitle

\section{Introduction} 
\label{sec:intro}

The Born-Oppenheimer (adiabatic) approximation \cite{BO1,BO2}, separating
the calculations of the electronic and nuclear wavefunction, is one of the fundamental
approximations in quantum chemistry.  It, however,  breaks down dramatically if two electronic
surfaces are nearly degenerate, i.e. the energy difference is within the vibrational energy splitting \cite{Horst84}.
In this case, the two adiabatic electronic states change their characters rapidly when the
nuclei move, and therefore non-adiabatic coupling is introduced.  
If the system has more than one nuclear degree of freedom, the two adiabatic
surfaces will often intersect each other. In other words, the system will have 
a conical intersection \cite{Horstbook}. The presence of a conical intersection
typically introduces new, dense spectral bands and hence changes the experimental
observations, e.g. the photoelectron spectrum, dramatically \cite{Horst84}.  For instance, there 
is a ``mysterious'' band found in the energy range from 9.5 to 9.9 eV in the butatriene photoelectron 
spectrum \cite{Brogli74}. This mysterious band can be explained via a vibronic coupling Hamiltonian, constructed 
from diabatic electronic basis and nuclear normal modes \cite{Lenz77}.  Similar situations can be found in
other systems as well, e.g. in molecules like allene, benzene, pyrazine, and $\text{SO}_2$ 
\cite{Mahapatra01,Doescher97,Raab99,Leveque13}.
Besides, the existence of conical intersections provides a fast non-radiative relaxation channel,
which can quench fluorescence \cite{Bearpark96} or introduce molecular isomerization \cite{Todd07}.
Nowadays, the vibronic coupling Hamiltonian together with nuclear dynamics calculations 
has become the standard treatment of non-adiabatic coupling systems \cite{Graham04}.  
For such a method, the total wavefunction ansatz is always written 
as \emph{ a sum of products of electronic and nuclear wavefunction over all involved electronic states} \cite{Horst84}. 

In contrast, 
there are attemps to go beyond the usual Born-Oppenheimer approximation by 
forcing an exact factorization on the total wavefunction. 
Namely, \emph{the total wavefunction is a single product of one electronic and one nuclear wavefunction} 
\cite{Hunter81,Gross05,Lenz13}.  In the literature a non-rotating diatomic system ($\text{H}_2^+$ and $\text{H}_2$) 
with only one vibrational mode (1D) was successfully studied \cite{Czub78,Gross05,Patrick06,Gross10}.
The most astonishing discovery from the 1D study is that such a wavefunction ansatz leads to
a ``spiky" potential and a nodeless nuclear wavefunction \cite{Hunter81,Czub78,Patrick06}.  
The only exception where the nuclear wavefunction can have a node is via symmetry, see Ref.~\cite{Gross05} 
for an example. Later studies focused on nuclear dynamics simulations with the 
exact factorized time-dependent potential of the 1D system \cite{Gross10}.  Till now, features related to conical 
intersections, which requires the presence of at least two nuclear degrees of freedom \cite{Todd10}, 
have never been studied with the exact factorized total wavefunction ansatz. 
In this paper, we will apply the single product wavefunction ansatz to a realistic
two-mode system, namely butatriene, and discuss the origin of the spiky potential
and its relation to the vibronic coupling effect.

\section{Theory}
\label{sec:theory}

Let us begin with introducing our system, which is a linear vibronic coupling model with diabatic
electronic basis functions $\{ \varphi_1(\symbf{q}) , \varphi_2(\symbf{q}) \}$. The Hamiltonian reads \cite{Horst84}
 \begin{align} 
 \label{eq:Hdia}
      \mkern-27mu \mathbf{H} =\left(-\frac{\hbar\omega_x}{2}\frac{\partial^2}{\partial Q_x^2}-\frac{\hbar\omega_y}{2}\frac{\partial^2}{\partial Q_y^2}     \right)\mathbf{1} +  \left( \begin{array}{cc}
      \!\!\! V_0+E_1+\kappa_1 Q_x & \lambda Q_y \!\!\! \\
      \!\!\! \lambda Q_y  & V_0+E_2+\kappa_2 Q_x \!\!\!
      \end{array} \right) 
 \end{align}
 where $V_0=\frac{\hbar\omega_x}{2}Q_x^2+\frac{\hbar\omega_y}{2}Q_y^2$.
 The normal modes $Q_x$ and $Q_y$, appearing in the diagonal and off-diagonal matrix elements, 
 are termed tuning mode and coupling mode, respectively.  The normal mode frequencies 
 $\omega_x$ and $\omega_y$, the energies of the diabatic states $E_1$ and $E_2$,
 and the coupling constants $\kappa_1$, $\kappa_2$, $\lambda$ can be obtained via 
 diabatizing the adiabatic potentials \cite{Horstbook}.
Diagonalizing the Hamiltonian yields
\begin{align}
\label{eq:eigensolution}
\mathbf{H} \left( \begin{array}{c} \chi_1^{(n)} \\ \chi_2^{(n)} \end{array} \right) = E_n \left( \begin{array}{c} \chi_1^{(n)} \\ \chi_2^{(n)} \end{array} \right) \;,
\end{align}
where $E_n$ is the $n$-th vibronic energy eigenvalue and $\{ \chi_1^{(n)}(\symbf{Q}) , \chi_2^{(n)} (\symbf{Q}) \}$ 
the $n$-th vibronic eigenfunction.  
The total wavefunction for each vibronic eigenfunction then reads
\begin{align}
\label{eq:totalwf}
\Psi_n(\symbf{q},\symbf{Q})=\varphi_1(\symbf{q})\chi_1^{(n)}(\symbf{Q}) + \varphi_2(\symbf{q})\chi_2^{(n)}(\symbf{Q})  \;,
\end{align}
where $\symbf{q}$ and $\symbf{Q}$ denote the electronic and nuclear degrees of freedom, respectively.
For the dynamics calculation, the total wavepacket is a linear combination of many $\Psi_n$. Here we
will concentrate on the individual eigenfunction and refer to $\Psi_n$ as our total wavefunction.

How to impose the single product condition on  $\Psi_n(\symbf{q},\symbf{Q})$?  
First, we can take a common part $\bar{\chi}_n$ out of $\chi_1^{(n)}$ and $\chi_2^{(n)}$
and regroup everything else as one single electronic wavefunction $\bar{\varphi}_n$.   
The $\bar{\chi}_n$ then represents the exact factorized nuclear wavefunction, or the exact nuclear
wavefunction for abbreviation.  Therefore, the wavefunction ansatz now reads  
\begin{align}
\label{eq:wfansatz}
\Psi_n(\symbf{q},\symbf{Q})=\left( \varphi_1(\symbf{q}) C_1^{(n)}(\symbf{Q}) + \varphi_2(\symbf{q}) C_2^{(n)} (\symbf{Q}) \right) \bar{\chi}_n \left( \symbf Q \right)  = \bar{\varphi}_n (\symbf{q},\symbf{Q}) \bar{\chi}_n (\symbf{Q})  \;,
\end{align}
where the coefficients $C_1^{(n)}$ and $C_2^{(n)}$ depend strongly on $\symbf{Q}$. 
The exact (factorized) electronic wavefunction $\bar{\varphi}_n$,
being a linear combination of diabatic electronic basis states $\{ \varphi_1 , \varphi_2 \}$,
consequently also depends strongly on $\symbf{Q}$.  The wavefunctions $\bar{\varphi}_n$ 
and $\bar{\chi}_n$ are all normalized: $\bar{\varphi}_n$ is normalized at each nuclear 
geometry $\symbf{Q}$ via integrating over all electronic degrees of freedom 
($\langle \bar{\varphi}_n | \bar{\varphi}_n \rangle_{\symbf{q}}$), while 
 $\bar{\chi}_n$ is normalized according to $\langle \bar{\chi}_n | \bar{\chi}_n \rangle_{\symbf{Q}}$.   
Still, the partitioning between $\bar{\varphi}_n$ and $\bar{\chi}_n$ in Eq.~\ref{eq:wfansatz} is not
unique. In other words, there are many ways to choose $\bar{\chi}_n$. 
Here we introduce one more condition on $\bar{\varphi}_n$ in order to achieve
a unique partition, namely, we require $\bar{\chi}_n$ to be real and positive so that 
$C_1^{(n)}$ and $C_2^{(n)}$ follow the sign of $\chi_1^{(n)}$ and $\chi_2^{(n)}$.  
This condition directly yields a nodeless $\bar{\chi}_n$, whose sign can never 
change in the whole nuclear space.  
Following the normalization condition of $\bar{\varphi}_n$ ($|C_1^{(n)}|^2+|C_2^{(n)}|^2=1$),
$C_1^{(n)}$ and $C_2^{(n)}$ can now be chosen as $\cos \theta$ and $\sin \theta$ with
$0 \leq \theta < 2\pi$, respectively.

Inserting the wavefunction ansatz, Eq.~\ref{eq:wfansatz}, and  the total Hamiltonian 
into the usual time-independent Schr\"odinger equation, we arrive at a coupled 
eigenvalue problem of the exact wavefunctions $\bar{\varphi}_n$ and $\bar{\chi}_n$. 
The working equations read \cite{Lenz13},
\begin{subequations}
\begin{align}
\label{eq:workingeq1}
& \bar{H}_{\text{el}}^{(n)} \bar{\varphi}_n = \bar{E}_{\text{el}}^{(n)} \bar{\varphi}_n   \\
\label{eq:workingeq2}
& \bar{H}_{\text{N}}^{(n)} \bar{\chi}_n = E_n \bar{\chi}_n   \;,
\end{align} 
\end{subequations}
where $\bar{H}_{\text{N}}^{(n)}=T_{\text{N}}  + \bar{E}_{\text{el}}^{(n)}$ is the exact nuclear Hamiltonian,
containing the nuclear kinetic energy operator $T_{\text{N}}$ and the exact potential $\bar{E}_{\text{el}}^{(n)}$,
while $\bar{H}_{\text{el}}^{(n)}$ is the exact electronic Hamiltonian. 
Interestingly, the exact electronic Hamiltonian $\bar{H}_{\text{el}}^{(n)}$ is different from
the usual electronic Hamiltonian $H_{\text{el}}$ and is given by \cite{Lenz13}
\begin{align}
\label{eq:Hel}
\bar{H}_{\text{el}}^{(n)} =H_{\text{el}}+T_{\text{N}}-\sum_{\alpha} \hbar \omega_{\alpha} \nabla_{\alpha}(\ln\bar{\chi}_n)\nabla_{\alpha}  \;,
\end{align}
where $T_{\text{N}}-\sum_{\alpha} \hbar \omega_{\alpha} \nabla_{\alpha}(\ln\bar{\chi}_n)\nabla_{\alpha}$ 
is responsible for the non-adiabatic coupling, and $\alpha$ is the index for the nuclei.  
According to Eq.~\ref{eq:Hel}, the nuclear motion
now couples directly to the electronic motion, and hence one has to solve 
for $\bar{\varphi}_n$ and $\bar{\chi}_n$ simultaneously.  To be more precise, one should use
an iterative procedure, where in each iteration one solves the eigenvalue problem of the 
Hamiltonians $\bar{H}_{\text{el}}^{(n)}$ and $\bar{H}_{\text{N}}^{(n)}$.  
Since $\bar{\chi}_n$ depends on $n$, the 
exact electronic Hamiltonian $\bar{H}_{\text{el}}^{(n)}$ and potential $\bar{E}_{\text{el}}^{(n)}$
are also $n$-dependent!   That is to say, there is one corresponding $\bar{\varphi}_n$ for 
each $\bar{\chi}_n$.  With such a wavefunction ansatz like Eq.~\ref{eq:wfansatz}, two 
different $\bar{\chi}$ cannot have the same $\bar{\varphi}$, and thus the usual picture
that one electronic state accommodates many different vibrational levels is no longer applicable.  
This is the price one pays for going beyond the Born-Oppenheimer approximation with
a single product wavefunction.  The advantage of this treatment is that the full correlation
between electrons and nuclei is considered simultaneously, i.e. the molecular vibration
is now also correlated with the electronic motion.

A straightforward simulation based on Eqs.~\ref{eq:workingeq1},\ref{eq:workingeq2} is
of course very expensive, but there is a shortcut for evaluating $\bar{E}_{\text{el}}^{(n)}$.
With the form of $\bar{\varphi}$ as a linear combination of diabatic basis states,
Eq.~\ref{eq:workingeq1} yields $\bar{E}_{\text{el}}^{(n)}$, which reads
\begin{align}
\label{eq:Hel2}
\bar{E}_{\text{el}}^{(n)} = \langle \bar{\varphi}_n | \bar{H}_{\text{el}}^{(n)} | \bar{\varphi}_n \rangle_{\symbf{q}} = \left( \begin{array}{cc} C_1^{(n)} & C_2^{(n)} \end{array} \right)  
\left( T_{\text{N}} \mathbf{1} + \mathbf{V}_{\text{dia}} \right) \left( \begin{array}{c} C_1^{(n)} \\ C_2^{(n)} \end{array}\right)  \;,
\end{align}
where $\mathbf{V}_{\text{dia}}$ denotes the diabatic potential matrix, which is given in Eq.~(\ref{eq:Hdia}).
Note that $\langle \bar{\varphi}_n | \nabla_{\alpha} \bar{\varphi}_n \rangle_{\symbf{q}}=0$.
One recalls that $C_1^{(n)}$ and $C_2^{(n)}$ are chosen as $\cos \theta$ and $\sin \theta$
with $0 \leq \theta < 2\pi$. Consequently, we know
\begin{align}
\label{eq:theta}
\tan \theta (\symbf{Q}) = \frac{\sin \theta}{\cos \theta}=\frac{C_2^{(n)}\bar{\chi}_n}{C_1^{(n)}\bar{\chi}_n}=\frac{\chi_2^{(n)}}{\chi_1^{(n)}}   \;.
\end{align} 
This equation states that $C_1^{(n)}$ and $C_2^{(n)}$ can be evaluated from the n-th
vibronic eigenfunction, and then one can construct $\bar{E}_{\text{el}}^{(n)}$ from the 
coefficients according to Eq.~\ref{eq:Hel2}.  The whole problem then reduces
to solving the nuclear eigenvalue problem as shown in Eq.~\ref{eq:workingeq2}.  
According to Eq.~\ref{eq:workingeq2}, diagonalizing $\bar{H}_{\text{N}}^{(n)}$
will again yield the energy eigenvalue $E_n$ and eigenfunction $\bar{\chi}_n$,
which can be compared with those obtained from the original diabatic Hamiltonian $\mathbf{H}$
of Eq.~\ref{eq:Hdia}.
We stress that this procedure is only for investigating features of $\bar{E}_{\text{el}}^{(n)}$
and $\bar{\chi}_n$, not for solving the full non-adiabatic coupling problem iteratively;
rather we need the eigenfunctions of the original non-adiabatic problem.

In our following calculation, an effective two-mode model for the butatriene system is taken as example,
with parameters listed in Tab.~\ref{tab:parameter}. The model is simple but 
sufficient to explain the experimental photoelectron spectrum \cite{Lenz77,Horst84}, 
and the result was also confirmed by a simulation with a full 18-mode MCTDH 
calculation \cite{Graham01}.  

\begin{table}
\begin{tabular}{|c|c|c|c|c|c|c|}
 \hline 
   $E_1$ & $E_2$ & $\omega_x$ & $\omega_y$ & $\kappa_1$ & $\kappa_2$ & $\lambda$    \\
 \hline
   9.45 & 9.85 & 0.2578 & 0.0913   &  -0.2121 &  0.2546 & -0.3182  \\ 
 \hline
\end{tabular}
\caption{Parameters of an effective two-mode model of butatriene, taken from Ref.~\cite{Lenz77}. The energy unit is eV. }
\label{tab:parameter}
\end{table}


\section{Results and Discussion}

\subsection{Non-adiabatic coupling with one vibrational mode}

\begin{figure}
\centering 
\hspace{-0.5cm}
\includegraphics[width=5.5cm]{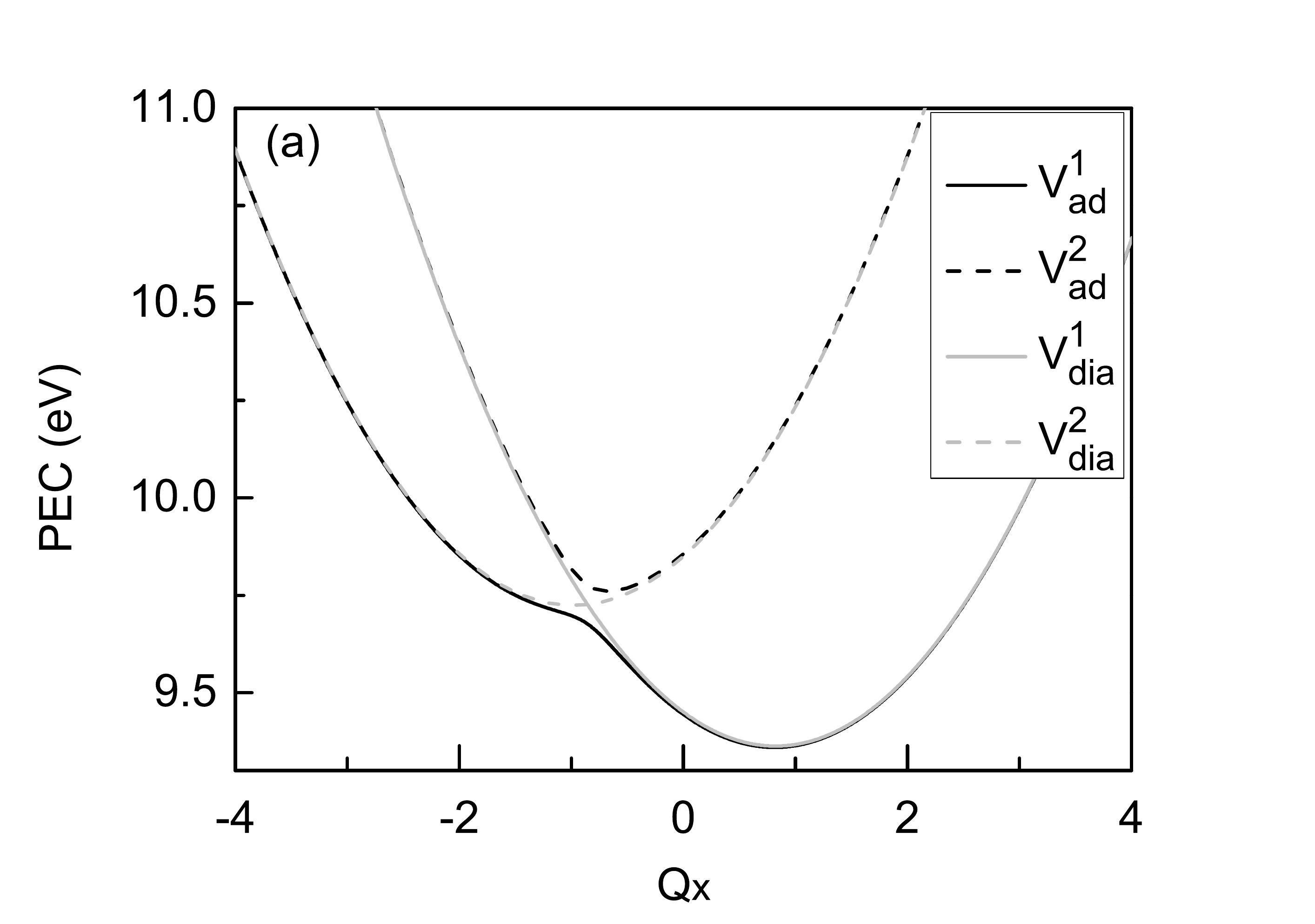} 
\hspace{-0.5cm}
\includegraphics[width=5.5cm]{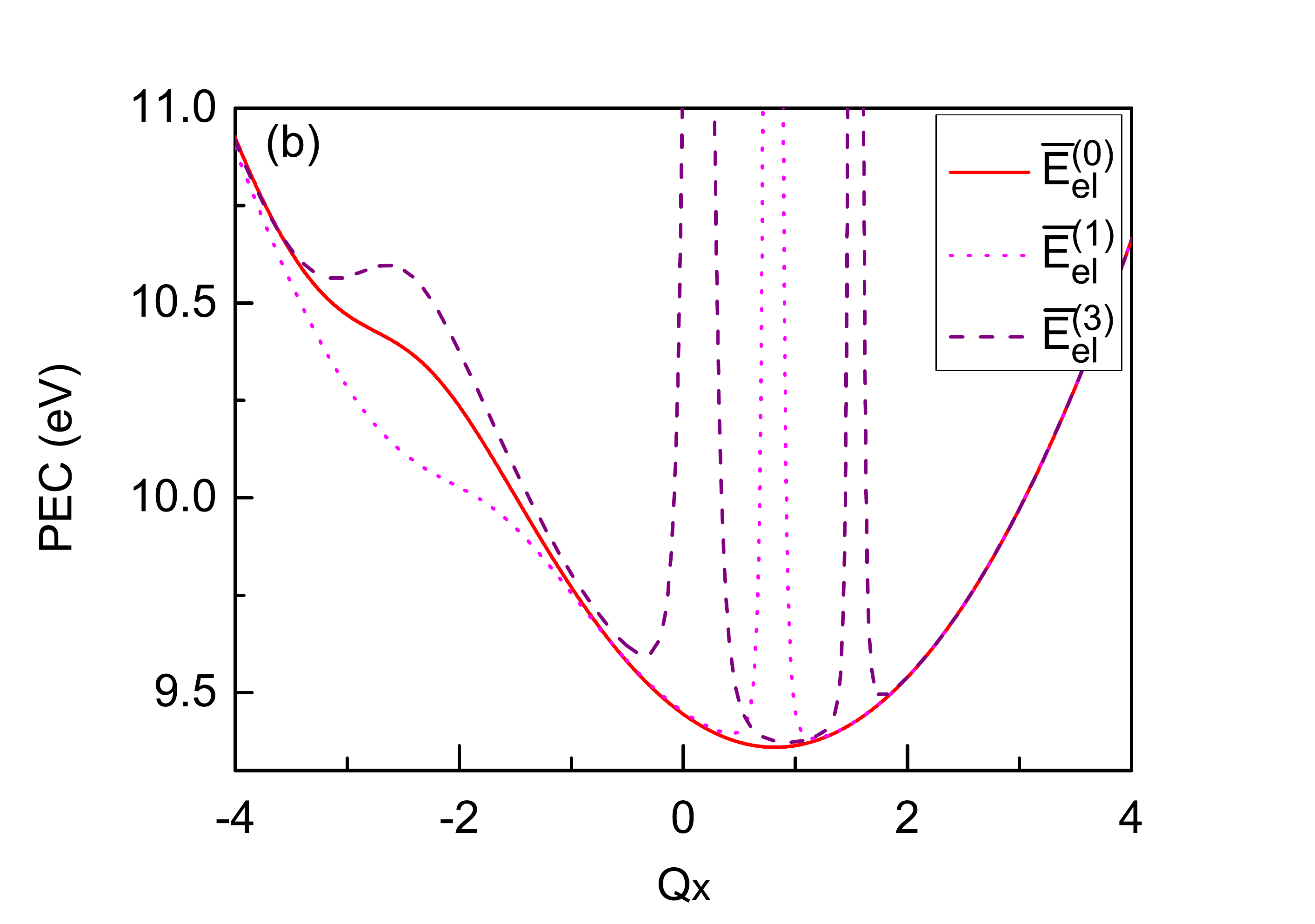} 
\hspace{-0.5cm}
\includegraphics[width=5.5cm]{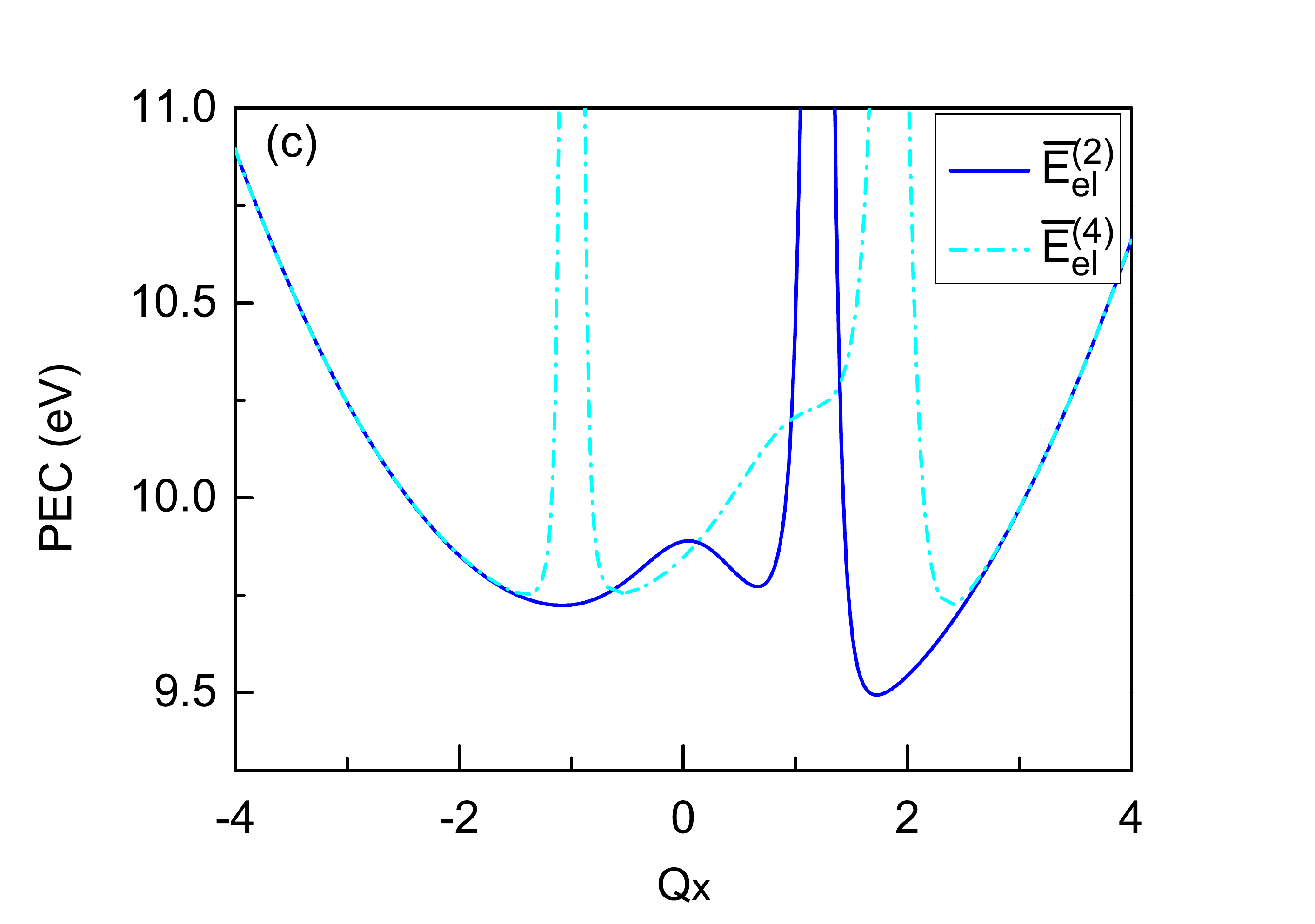} 
\caption{(Color online) Potentials. The adiabatic and diabatic potentials
are depicted in panel (a), while the exact factorized potentials $\bar{E}_{\text{el}}$
are depicted in panel (b) and (c).  The exact potentials form two different
groups, and each group follows strongly a diabatic potential, e.g. the $\bar{E}_{\text{el}}$
depicted in panel (b) follow strongly $\text{V}_{\text{dia}}^1$ from panel (a).
Additionally, $\bar{E}_{\text{el}}$ in general has strong barriers, which
leads to the node-avoiding feature of $\bar{\chi}$. See also Fig.~\ref{fig:chi1d}. }
\label{fig:pot1d}
\end{figure}

To make the physics transparent, we first consider only the tuning mode $Q_x$ and a 
constant coupling $\lambda$ in Eq.~\ref{eq:Hdia}. The coupling constant here is chosen to be
$0.05$ eV to show a typical weakly avoided crossing, while the other parameters are as listed 
in Tab.~\ref{tab:parameter}.  The adiabatic potentials, depicted in Fig.~\ref{fig:pot1d}~(a), have
the avoided crossing around 9.75 eV, implying a strong non-adiabatic effect. Otherwise, the
adiabatic potentials follow the diabatic potentials well.  On the other hand, 
the exact $\bar{E}_{\text{el}} ^{(n)}$, depicted in panels (b) and (c), are divided into two
groups.  The group shown in panel (b) basically follows the \emph{diabatic} potential 
$V_{\text{dia}}^1$ with a lower minimum, while the other group, shown in panel (c), follows the \emph{diabatic} potential
$V_{\text{dia}}^2$.  As already discovered in Refs~\cite{Czub78,Hunter81,Gross05,Patrick06}, 
all $\bar{E}_{\text{el}} ^{(n)}$ have spikes, except for $\bar{E}_{\text{el}} ^{(0)}$. 
These spikes actually come from the kinetic energy operator applied on the eletronic wavefunction
$\bar{\varphi}_n$, i.e. $\langle \bar{\varphi}_n | \hat{T}_{\text{N}} | \bar{\varphi}_n \rangle$,
which is closely related to how the non-adiabatic coupling originates. In fact, replacing 
$\bar{\varphi}_n$ by adiabatic wavefunctions, this expectation value would yield the diagonal correction term
automatically.  The proof of a spiky potential is simple.  Replacing $C_1$ and $C_2$ by 
$\cos \theta$ and $\sin \theta$, the kinetic energy operator contribution to $\bar{E}_{\text{el}}$ in Eq.~\ref{eq:Hel2}
reads (omitting $(n)$ for simplicity),
\begin{align}
\label{eq:keocon1d}
\left( \begin{array}{cc} C_1 & C_2 \end{array} \right)  T_{\text{N}} \mathbf{1} \left( \begin{array}{c} C_1 \\ C_2 
\end{array}\right)  
= \cos \theta ( -\frac{\hbar \omega_x}{2} \frac{\text{d}^2}{ \text{d} Q_x^2} \cos \theta) + \sin \theta ( -\frac{\hbar \omega_x}{2} \frac{\text{d}^2}{ \text{d} Q_x^2} \sin \theta) 
= \frac{\hbar \omega_x}{2} \left( \frac{\text{d} \theta}{\text{d} Q_x} \right) ^2 \;.
\end{align}
When $\chi_1$ and $\chi_2$ have a node, $C_1$ and $C_2$
actually change sign by construction. This then leads 
to a rapid variation in $\theta$. For example, if $\chi_1$ or $\chi_2$
has a node,  $\theta$ moves rapidly from one quadrant to the other.
If both $\chi_1$ and $\chi_2$ have nodes in a small range of $Q_x$, 
$\theta$ changes by two quadrants within this range.  If both $\chi_1$ and $\chi_2$ 
have a node at the same $Q_x$, $\theta$ must jump by $\pi$.  
Consequently, the derivative square $\left( \frac{\text{d} \theta}{\text{d} Q_x} \right) ^2$ 
will behave like a $\delta$-function and therefore cause a spike. 
As for the expectation value of the diabatic potential 
$\langle \bar{\varphi}_n | \mathbf{V}_{\text{dia}} | \bar{\varphi}_n \rangle$,
it forms the basic shape of the potential $\bar{E}_{\text{el}}^{(n)}$, i.e. all other parts except the spikes.

\begin{table}
\begin{tabular}{|c|c|c|c|c|c|c|}
 \hline 
   $n$ & $H$ & $\bar{H}_{\text{N}}^{(n)}$  & $H $[$\lambda=0$]  &  $H_{\text{ad}}$  & $H_{\text{B.-H.}}$  & $\bar{H}_{\text{N}}^{(0)}$   \\
 \hline
   0 & 9.4878   & 9.4878  & 9.4916 & 9.4857    &  9.4938 &  9.4878  \\ 
   1 & 9.7404   & 9.7404   & 9.7494 & 9.7243    &  9.7686 &  9.7428 \\
   2 & 9.8561   & 9.8561    & 9.8532  & 9.9205   &  9.9762 &   -- \\
   3 & 10.0087 & 10.0087  & 10.0071  & 9.9527 & 10.0517 &  9.9941 \\
   4 & 10.1088 & 10.1088 & 10.1109  & 10.1132 & 10.1230 &   -- \\
   5 & 10.2656 & 10.2656  & 10.2649  & 10.3054 & 10.3672 &  10.2361\\
   6 & 10.3693 & 10.3693  & 10.3687  & 10.3247 & 10.3747 &  -- \\
   7 & 10.5205 & 10.5205  & 10.5226  & 10.5346 & 10.5435 &   10.4560 \\
   8 & 10.6290 & 10.6290  & 10.6265  & 10.6339 & 10.6500 &    -- \\
   9 & 10.7781 & 10.7781 & 10.7804  & 10.7520 & 10.7851 &  10.6458 \\
 \hline
\end{tabular}
\caption{Energy eigenvalues obtained from different Hamiltonians. The unit is eV. 
Note that $\bar{H}_{\text{N}}^{(0)}$ only produces approximations to the states with the 
lower-minimum diabatic electronic potential $V_{\text{dia}}^1$, i.e. $n=0,1,3,5,\cdots$. For $n=2,4,6,8$,
the eigenfuctions are dominated by another diabatic potential, $V_{\text{dia}}^2$.
For energies larger than 10.25 eV, the potential $\bar{E}_{\text{el}}^{(0)}$ deviates strongly from
the diabatic potential $V_{\text{dia}}^1$, and hence its eigenvalues gets becomes than the
adiabatic ones.  }
\label{tab:E1d}
\end{table}    

\begin{table}
\begin{tabular}{|c|c|c|c|c|c|}
 \hline 
   $n$ &  $\bar{H}_{\text{N}}^{(n)}$  & $H $[$\lambda=0$]  &  $H_{\text{ad}}$  & $H_{\text{B.-H.}}$  & $\bar{H}_{\text{N}}^{(0)}$   \\
 \hline
   0 &  1.0000 & 0.9965 & 0.9937 &  0.9948 &  1.0000 \\ 
   1 &  1.0000 & 0.9679 & 0.9621 &  0.9684 &  0.9904 \\
   2 &  1.0000 & 0.9521 & 0.7117  &  0.7283 &   -- \\
   3 &  1.0000 & 0.9819  & 0.4422 & 0.3824 &  0.9474 \\
   4 &  1.0000 & 0.9862 & 0.5657 & 0.5137  &   -- \\
   5 &  1.0000 & 0.9852 & 0.3735 & 0.4296  &  0.9038 \\
   6 &  1.0000 & 0.9904 & 0.4553 & 0.4566  &  -- \\
   7 &  1.0000 & 0.9839 & 0.7328 & 0.7043  &  0.7956 \\
   8 &  1.0000 & 0.9849 & 0.6283 & 0.6499  &    -- \\
   9 &  1.0000 & 0.9855 & 0.6941 & 0.7339  &  0.5825 \\
 \hline
\end{tabular}
\caption{Overlap between vibronic eigenfunctions, obtained via diagonalizing $H$, 
and eigenfunctions of different Hamiltonians.  Eigenfunctions are first transformed
to the diabatic basis in the overlap procedure.}
\label{tab:F1d}
\end{table}    

Why are the exact potentials $\bar{E}_{\text{el}}^{(n)}$ so similar to 
the diabatic potentials?  This phenomenon suggests that the diabatic 
electronic basis could be better than the adiabatic one in the current example. 
To confirm this idea, we compare eigenvalues obtained from different Hamiltonians.
In Tab.~\ref{tab:E1d}, the sorted exact energy eigenvalues obtained via diagonalizing
$\mathbf{H}$ are given, and the eigenvalues obtained from $\bar{H}_{\text{N}}^{(n)}$
indeed are identical to them. However, to achieve full convergence in solving 
Eq.~\ref{eq:workingeq2}, we have to use a sine-DVR \cite{Beck00} with 3200 points! 
This unusually large DVR size is needed for smoothly reproducing the spiky potentials 
$\bar{E}_{\text{el}}^{(n)}$ shown in Fig.~\ref{fig:pot1d}.  With a smooth but spiky 
potential, only the ground vibrational eigenfunction of $\bar{H}_{\text{N}}^{(n)}$
yields the exact $\bar{\chi}_n$.  This is due to the condition imposed on $\bar{\chi}_n$
that it must not change sign for the complete $Q$ space.
Next we look at the eigenvalues of $H$[$\lambda=0$], which are indeed close to 
the exact ones since the weak off-diagonal coupling $\lambda$ is like a small perturbation to the Hamiltonian. 
In contrast, the adiabatic approximation and Born-Huang approximation (adiabatic approximation
plus the diagonal correction) do not yield as good energy estimates as $H$[$\lambda=0$].
The overlap of these eigenfunctions with the exact eigenfunctions of $H$, see Tab.~\ref{tab:F1d},
also indicates that these two approximations are valid for only two states ($n$=0,1).
Unlike the adiabatic approximation, the overlaps listed in column $H$[$\lambda=0$] 
are almost one.  It is clear that the diabatic basis is indeed a best choice in this
example.  In fact, if a potential follows closely the diabatic potential, it will also
be a better choice than the adiabatic one. For instance, the eigenvalues and 
eigenfunctions of $\bar{H}_{\text{N}}^{(0)}$ are better than those obtained from 
the adiabatic approximation, as shown in the last columns of Tab.~\ref{tab:E1d} and Tab.~\ref{tab:F1d}.

\begin{figure}
\centering 
\includegraphics[width=8cm]{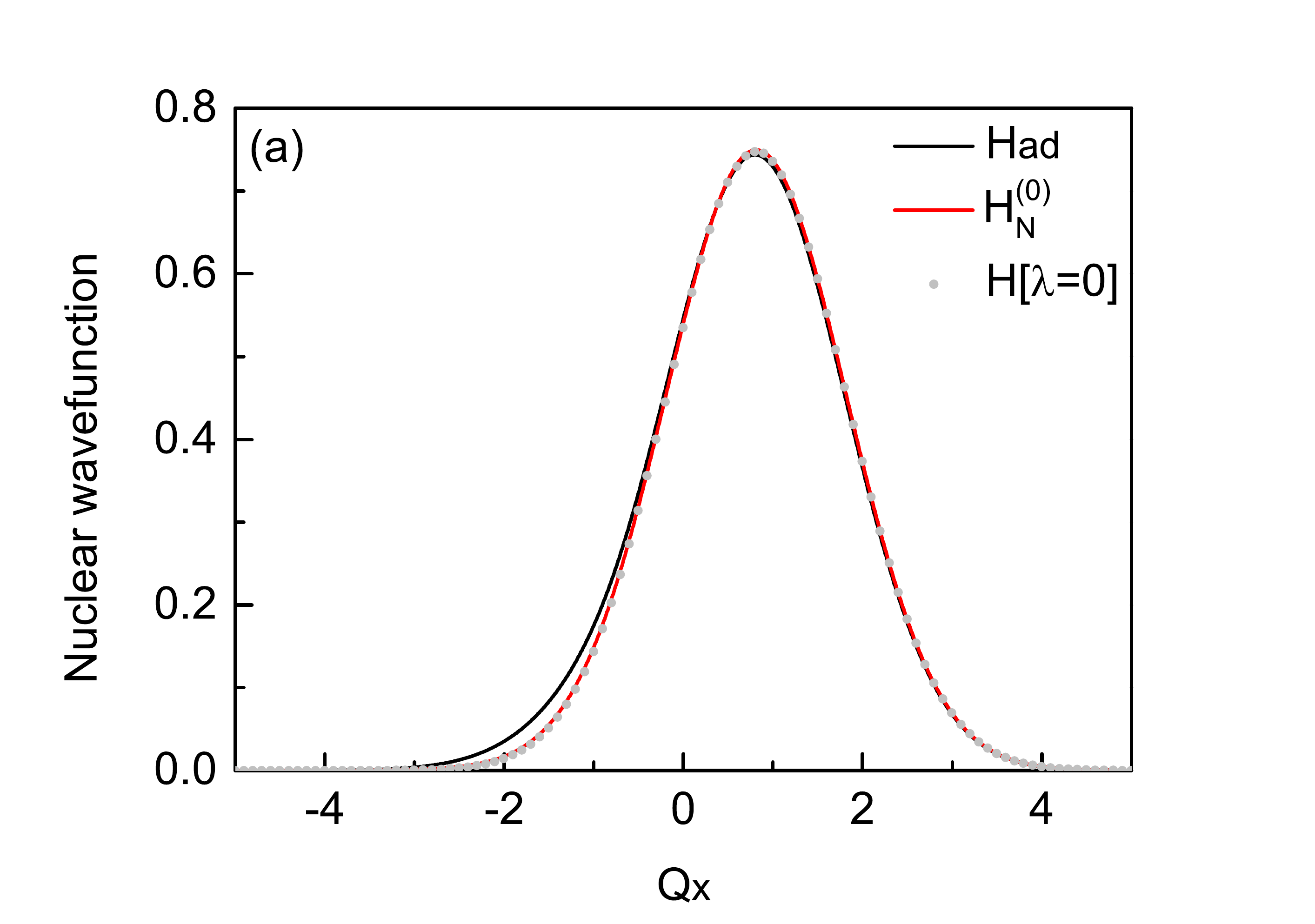} 
\includegraphics[width=8cm]{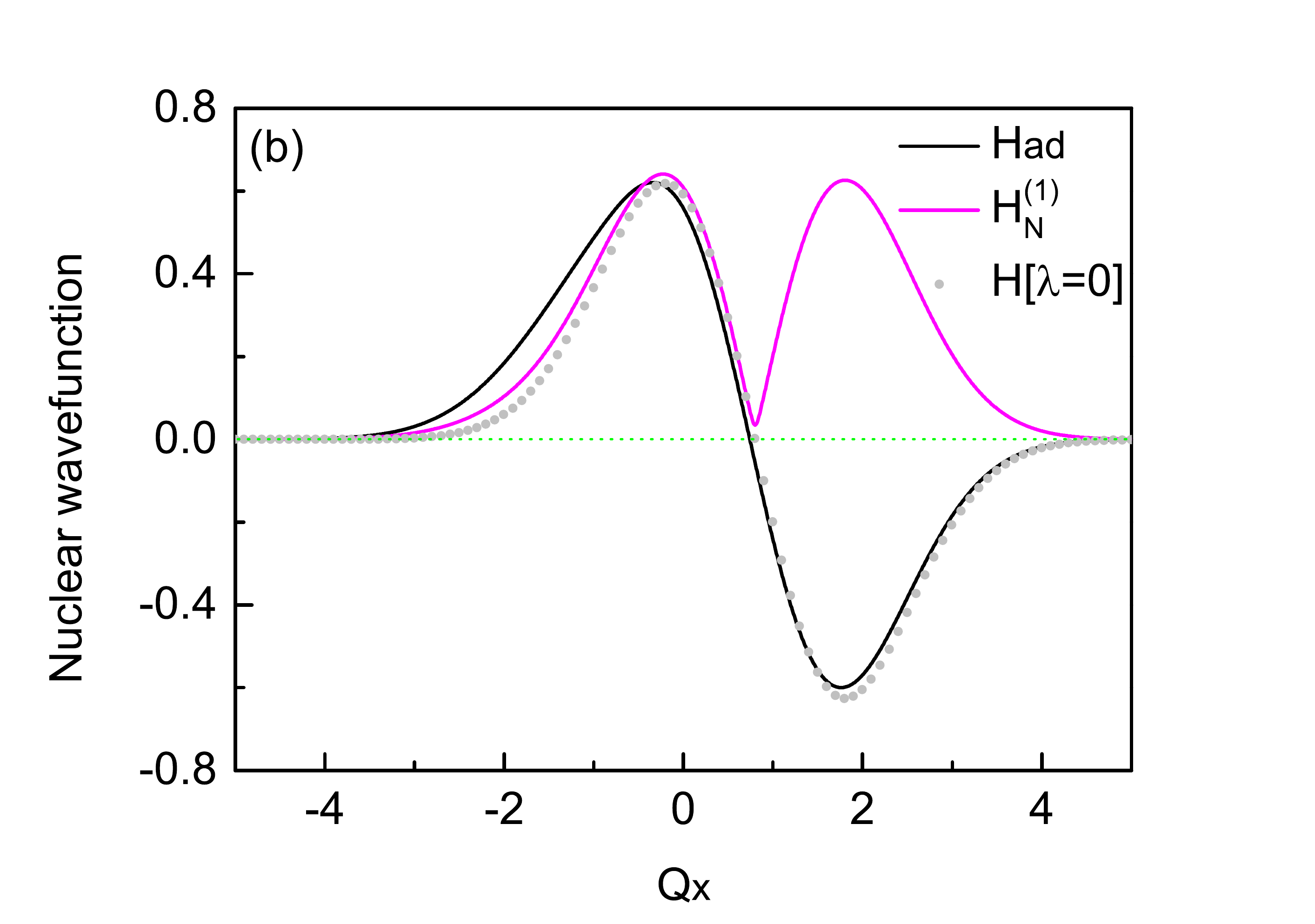} 
\includegraphics[width=8cm]{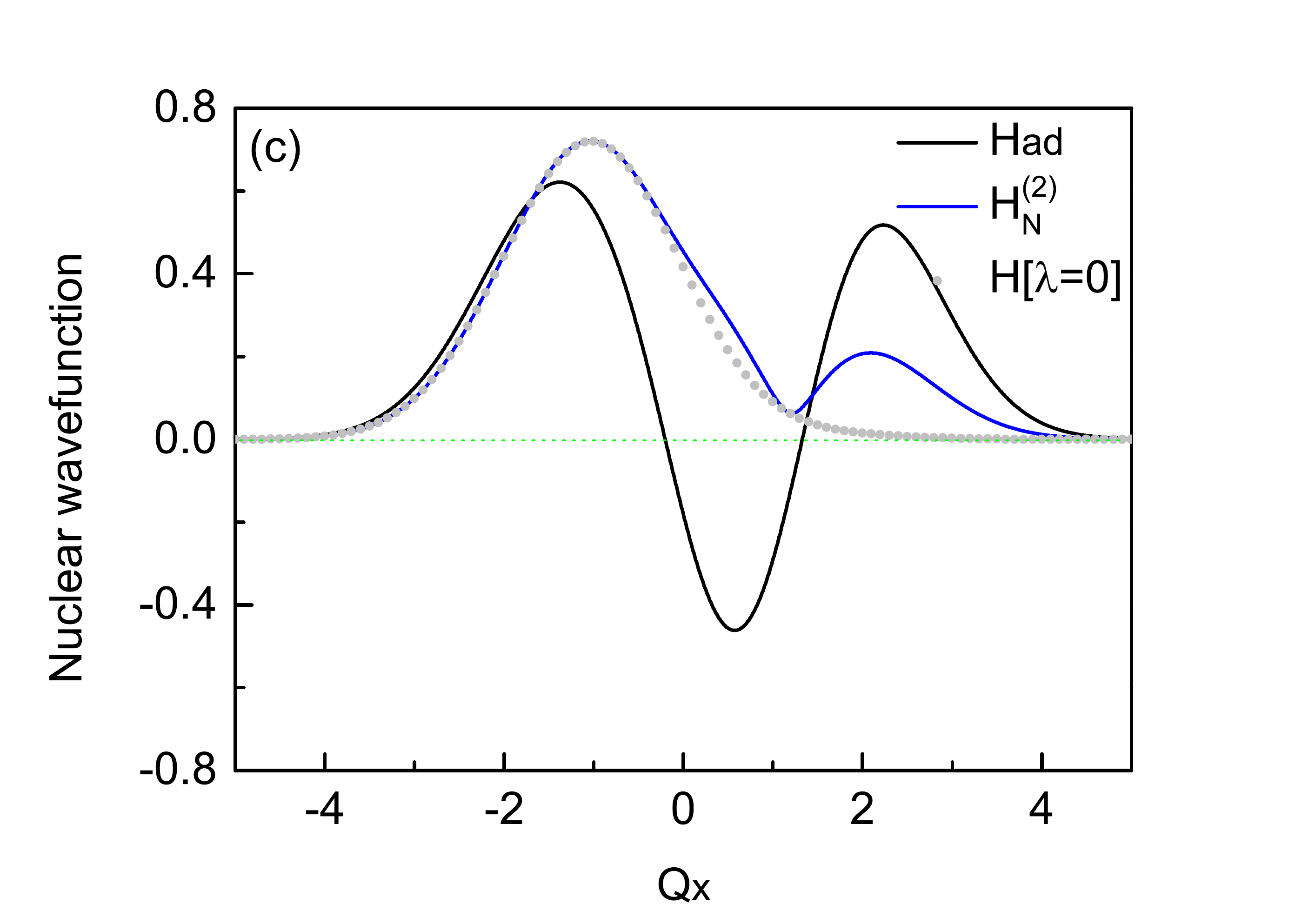} 
\includegraphics[width=8cm]{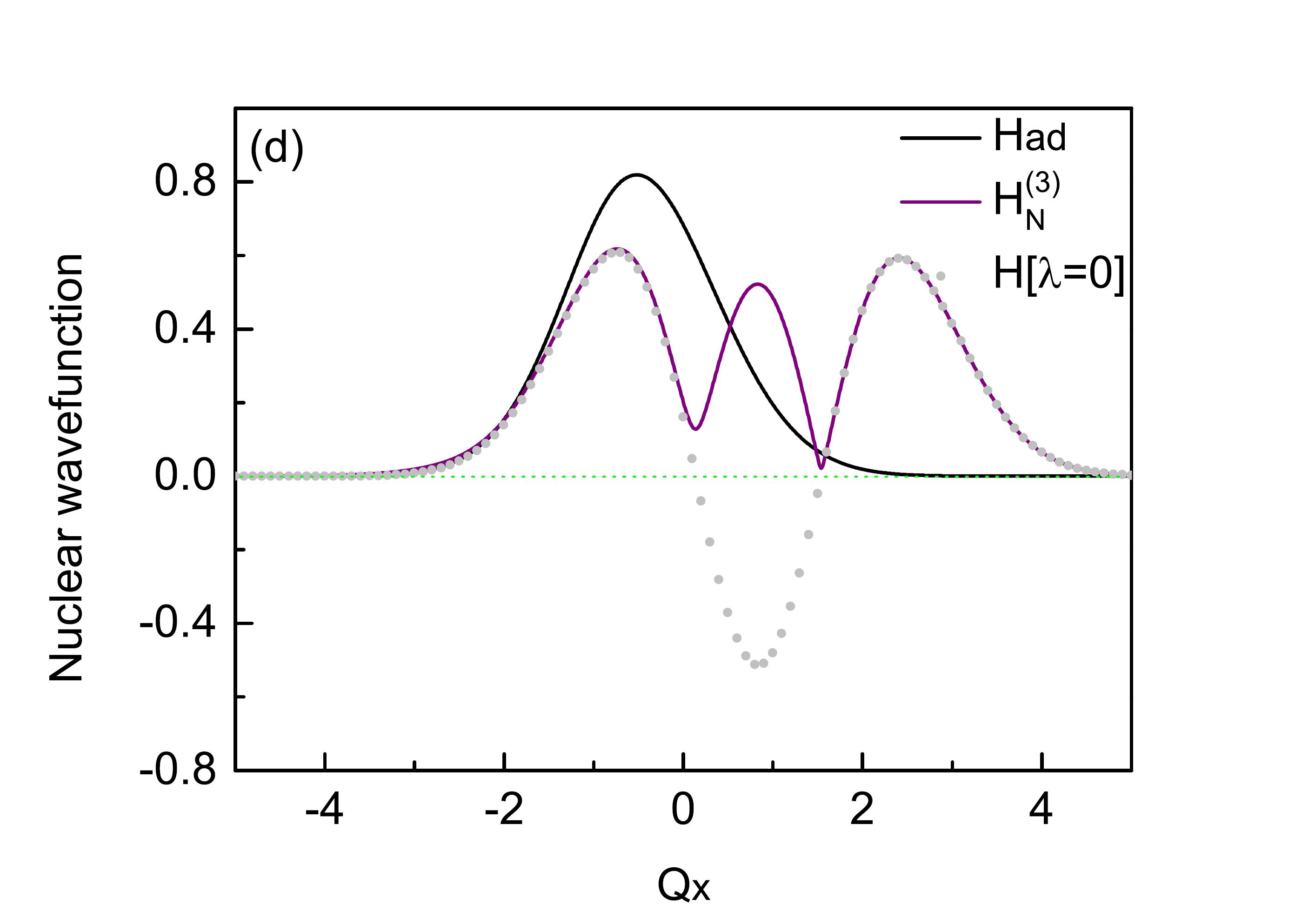} 
\caption{(Color online) Eigenfunctions for $n=0-3$. Black curves, colored curves,
gray dots are used to depict eigenfunctions of $H_{\text{ad}}$, 
$\bar{H}_{\text{N}}^{(n)}$, and $H$[$\lambda=0$], respectively. 
Panel (a): eigenfunctions for $n=0$. The adiabatic approximation 
is still valid for this state, but it then deviates from the nodeless $\bar{\chi}_n$
when $n=1$, as shown in panel (b).  
Panel (c) and (d) : eigenfunctions for $n=2$ and $n=3$, respectively.  
The $\bar{\chi}_2$ in panel (c) has a small tail, which is a fingerprint
of the non-adiabatic coupling. When comparing the adiabatic solutions and the diabatic solutions,
one finds that the energy order and wavefunction-based order (see text)
interchange.  For example, the adiabatic eigenfunction for $n=3$, shown 
by the black curve in panel (d), has a similar shape to $\bar{\chi}_2$ in panel (c).
This ordering-interchange phenomenon makes a meaningful comparison 
between adiabatic and exact solutions difficult.}
\label{fig:chi1d}
\end{figure}

Finally, how do the exact factorized wavefunctions $\bar{\chi}$ look like?
In Fig.~\ref{fig:chi1d} the $\bar{\chi}_n$ for $n=0-3$ are depicted in color lines, 
together with the corresponding adiabatic eigenfunctions (black lines) and the eigenfunctions
of $H$[$\lambda=0$] (gray dots), which is a good approximation to the exact vibronic eigenfunction
and will be termed the diabatic eigenfunction.  Shown in panel (a) are the eigenfunctions 
for $n=0$, and all of them are quite close to each other, except for the adiabatic one
which deviates a little.  
Starting from the next state $n=1$, see panel (b), $\bar{\chi}_1$
already has no nodes! This node-avoiding effect, as mentioned before, is
expected because we impose the no-sign-change condition on  $\bar{\chi}_n$.
Consequently $\bar{\chi}_n$ always appears as the ground state of $\bar{H}_{\text{N}}^{(n)}$.
In contrast to $\bar{\chi}_1$, the eigenfunctions of $H_{\text{ad}}$ and of $H$[$\lambda=0$]
both have a node.  In comparison, the eigenfunction of $H$[$\lambda=0$] is still
better than the adiabatic approximation, since the former coincides better with the exact $\bar{\chi}_1$ 
up to the "node avoiding" position.  There is another problem with the adiabatic approximation,
namely how to sort the eigensolutions. 
The current list is based on the usual ascending energy order. Yet if we look at the
$m$-th order adiabatic eigenfunction, it might have a similar shape as the $n$-th
order diabatic eigenfunction, where $ n \neq m$.  Hence, the wavefunction-based order and the energy-based
order interchanges when one compares adiabatic and diabatic solutions.
For example, the adiabatic eigenfunction for $n=3$, depicted by the black line in panel (d),
has a similar shape to the exact $\bar{\chi}_2$, shown in panel (c), and to the diabatic one as well. 
This eigenfunction, $\bar{\chi}_2$, is actually dominated by the ground state of the Hamiltonian 
with the diabatic potential $V_{\text{dia}}^2$, except for the small peak in $\bar{\chi}_2$, which
completely results from the non-adiabatic coupling. On the other hand, the adiabatic eigenfunction
for $n=2$, according to energy order, has a similar shape to the exact  $\bar{\chi}_3$ and the diabatic eigenfunction
shown in panel (d).  The wavefunction ordering and energy ordering clearly interchange in the
adiabatic solutions.   Based on the energy order, the overlap of adiabatic eigenfunctions with
the diabatic eigenfunctions is of course bad.
When one moves on to higher excited states, a meaningful comparison 
becomes more and more difficult.  We will encounter the
same problem again when discussing the two-mode butatriene example.

\subsection{Non-adiabatic coupling with two vibrational modes}

\begin{figure}
\centering 
\includegraphics[width=8cm]{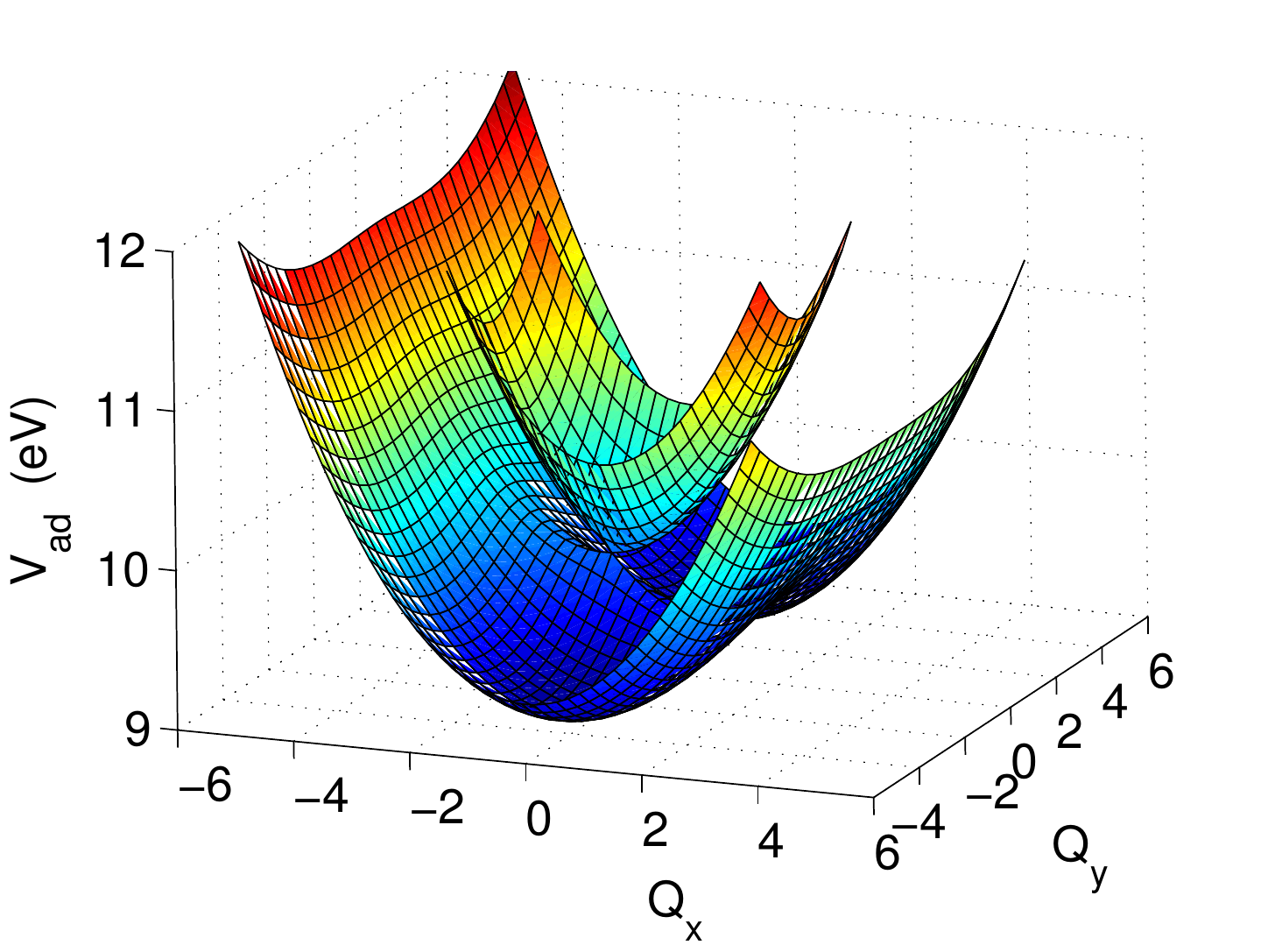} 
\caption{(Color online) Adiabatic potentials of butatriene with two effective modes. 
The two potential surfaces intersect at energy 9.7265 eV with $(Q_x,Q_y)=(-0.8571,0)$.}
\label{fig:Vad2d}
\end{figure}

Now we proceed to the realistic two-mode example of butatriene, where
the adiabatic potentials display a conical intersection.    
As shown in Fig.~\ref{fig:Vad2d}, the conical intersection occurs at 
energy 9.73 eV with $(Q_x,Q_y)=(-0.86,0)$.   The lower surface forms 
a double well potential,  showing the symmetry breaking effect due to the presence 
of the coupling mode $Q_y$ in the Hamiltonian \cite{Horst84,Fulton61,Fulton64}. 
Consequently, the lower eigenvalues are almost doubly degenerate. 

Let us first look at the adiabatic eigenvalues. For $H_{\text{ad}}$ and $H_{\text{B.-H.}}$
the eigenvalues shown in Tab.~\ref{tab:E2d} agree well with the exact ones. 
The eigenfunctions, on the other hand, do not behave as nice. To show this problem, 
the eigenfunction of $H_{\text{ad}}$ is labeled by $m$
and is overlapped with $n$-th exact vibronic eigenfunction of $H$ with $n$=0-50,
which corresponds to an energy range from 9.2381 to 10.0992 eV.  If the maximum
value of the overlap appears at the diagonal of the overlap matrix ($n=m$),
it means that the energy-based ordering agrees with the wavefunction-based
ordering. 
\begin{table}
\begin{tabular}{|c|c|c|c|c|c|}
 \hline 
   $n$ & $H$ & $\bar{H}_{\text{N}}^{(n)}$ & $H_{\text{ad}}$ & $H_{\text{B.-H.}}$ & $\bar{H}_{\text{N}}^{(0)}$  \\
 \hline
   0 & 9.2381 & 9.2381 & 9.2367 & 9.2383 & 9.2381 \\ 
   1 & 9.2381 & 9.2381 & 9.2367 & 9.2383 & 9.2381 \\
   2 & 9.3251 & 9.3251 & 9.3232 & 9.3254 & 9.3251 \\
   3 & 9.3253 & 9.3253 & 9.3235 & 9.3257 & 9.3254 \\
   4 & 9.4084 & 9.4084 & 9.4053 & 9.4091 & 9.4086 \\
   5 & 9.4113 & 9.4113  & 9.4091 & 9.4120 & 9.4116 \\
   6 & 9.4703 & 9.4703 & 9.4693 & 9.4709 & 9.4710 \\
   7 & 9.4704 & 9.4704 & 9.4694 & 9.4710 & 9.4711 \\
 \hline
\end{tabular}
\caption{First eight energy eigenvalues of different Hamiltonians for the butatriene example.
The energy from the exact diabatic Hamiltonian $H$ and the exact factorized Hamiltonian
$\bar{H}_{\text{N}}^{(n)}$ are identical. $\bar{H}_{\text{N}}^{(0)}$ yields
better energy eigenvalues for $n=0-5$ than the adiabatic approximation and Born-Huang approximation. }
\label{tab:E2d}
\end{table}   

\begin{figure}
\centering 
\includegraphics[width=8cm]{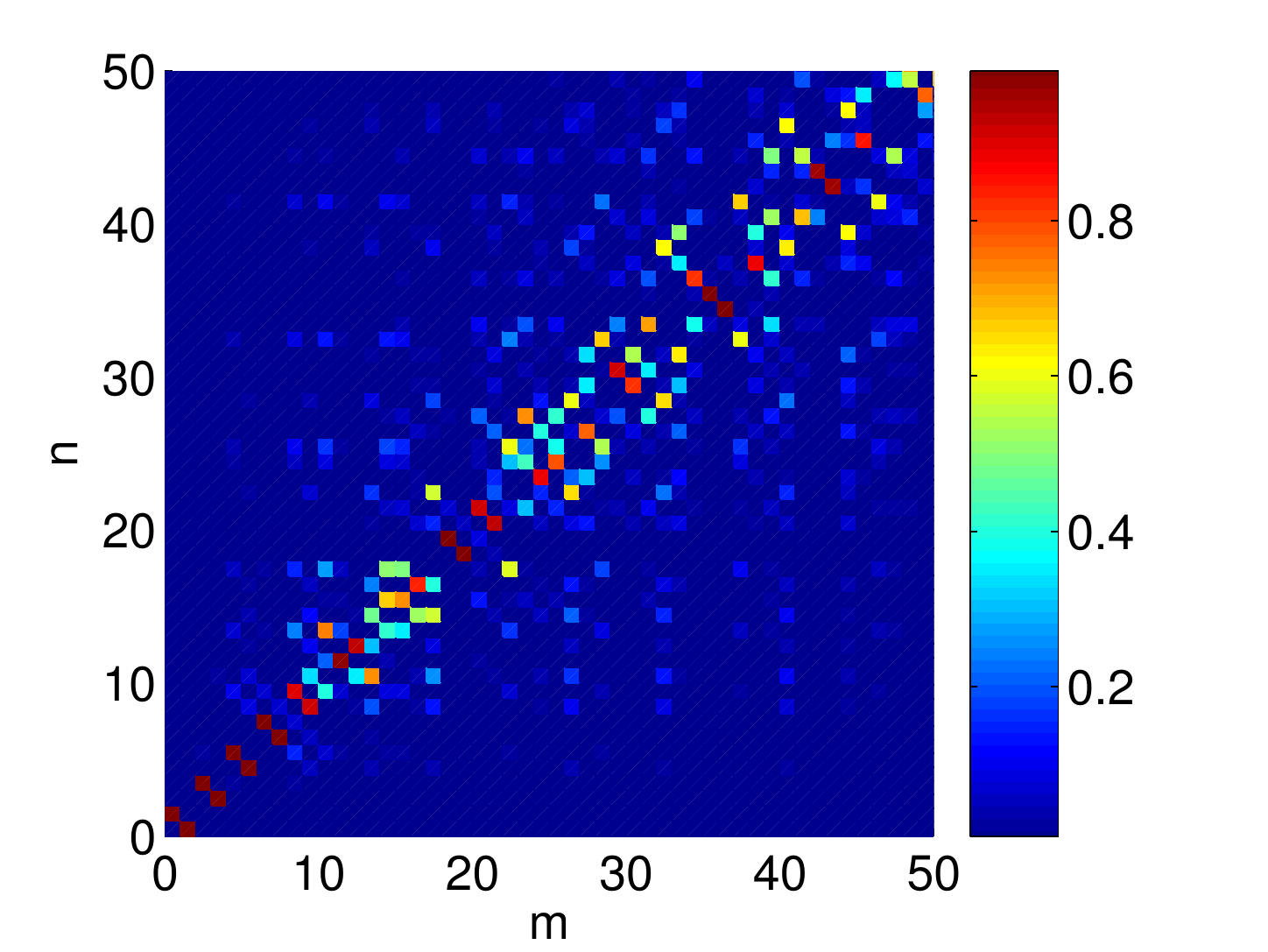} 
\caption{(Color online) Matrix of overlaps between the eigenfunctions of $H_{\text{ad}}$ ($m$) and $H$ ($n$).
For $m \leq 10$, the maximum overlap value only appears at the superdiagonal and the subdiagonal
of the matrix. This shows that the eigenfunction-based order interchanges with
the usual ascending energy order when one compares the adiabatic eigenfunctions to  the
vibronic eigenfunctions.  For  $m > 10$, the overlap values drop dramatically, which
shows that the adiabatic approximation fails for $m>10$, i.e. energies larger than 9.5 eV. }
\label{fig:olad}
\end{figure}  
As shown in Fig.~\ref{fig:olad}, the adiabatic eigenfunctions,
having a different order than the ascending energy order, 
nevertheless agree perfectly with the exact vibronic eigenfunctions for $m \leq 10$,
i.e. the overlap is nearly 1.
As for larger $m$, there is no one-to-one mapping between 
adiabatic eigenfunctions and the exact ones. The adiabatic approximation
indeed breaks down for energies larger than 9.5 eV \cite{Horst84}.
Additionally, $\bar{H}_{\text{N}}^{(0)}$ is also better than the adiabatic approximation,
see the data in Tab.~\ref{tab:E2d}. We also find that the overlaps between its eigenfunctions 
and the vibronic eigenfunctions to be nearly 1 for the first eight states, and the eigenfunction-based
ordering is the same as the energy-based ordering when one compares eigensolutions of 
$\bar{H}_{\text{N}}^{(0)}$ to the vibronic solutions.   
As a remark, we mention that the energy eigenvalues and eigenfunctions
of $\bar{H}_{\text{N}}^{(n)}$ are exact. For instance, eigenvalues in Tab.~\ref{tab:E2d}
and the corresponding eigenfunctions are numerically converged.  
For large $n$, this requires extremely fine grid spacing, e.g. $10^{-5}$, to obtain converged 
$\bar{\chi}_n$ from Eq.~\ref{eq:workingeq2}.  Hence we introduce an estimated  
$\bar{\chi}_n$ for $n \geq 8$ according to $\bar{\chi}_n=\frac{\chi_1^{(n)}}{\cos \theta}=\frac{\chi_2^{(n)}}{\sin \theta}$.
In this case, one can first evaluate $\theta$ from the vibronic eigenfunctions and then evaluate $\bar{\chi}$
according to the obtained $\theta$, but using a grid spacing $0.01$.  This allows us
to investigate the properties of $\bar{\chi}_n$ for $n$ up to $n=60$.
All eigenfunctions and eigenvalues of $H$, $H_{\text{ad}}$, and $H_{\text{B.-H.}}$ are numerically
converged up to $n=60$.
 
\begin{figure}
\centering 
\subfigure[]{
   \includegraphics[width=6cm]{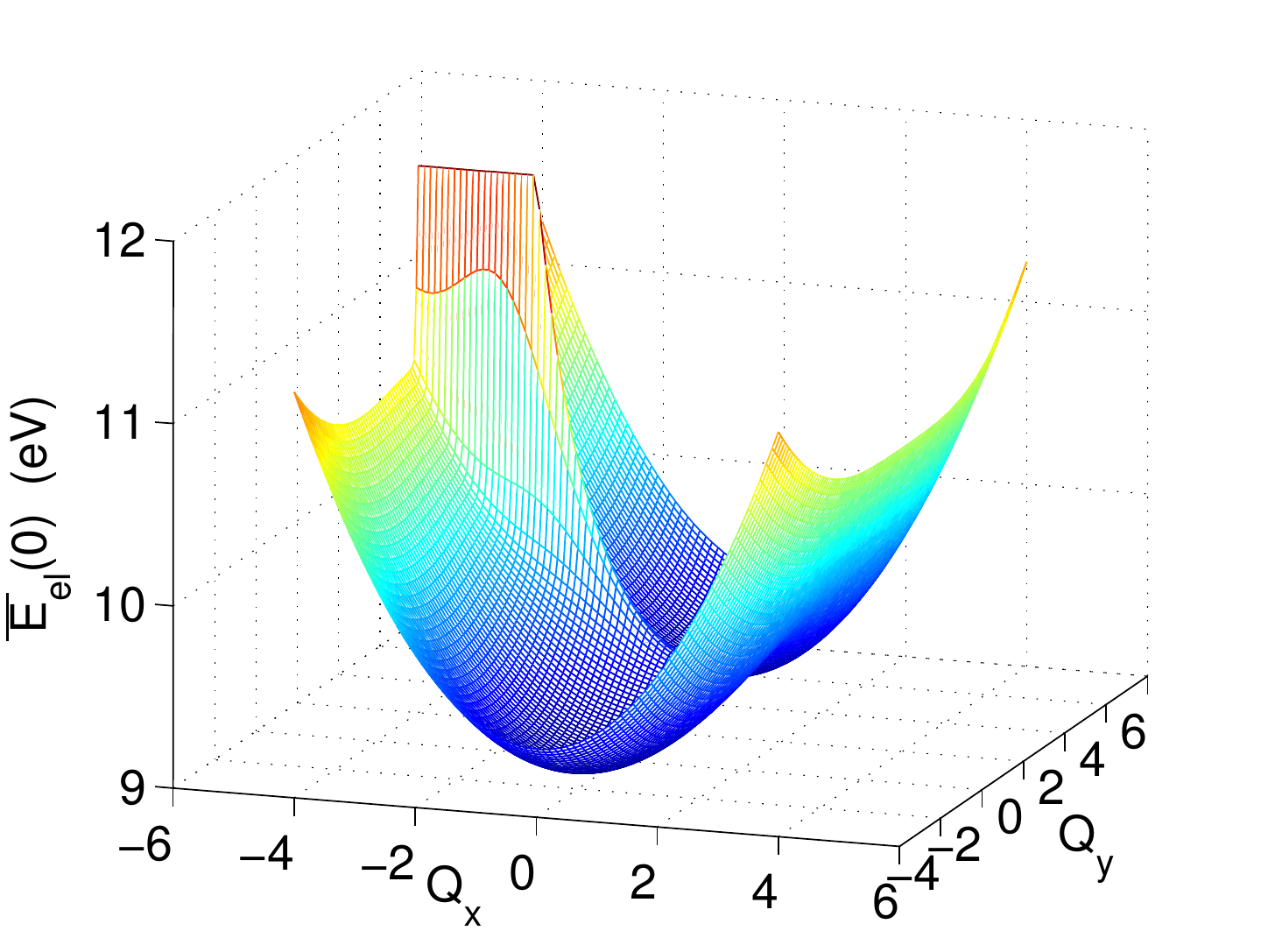}
 }
\subfigure[]{
   \includegraphics[width=6cm]{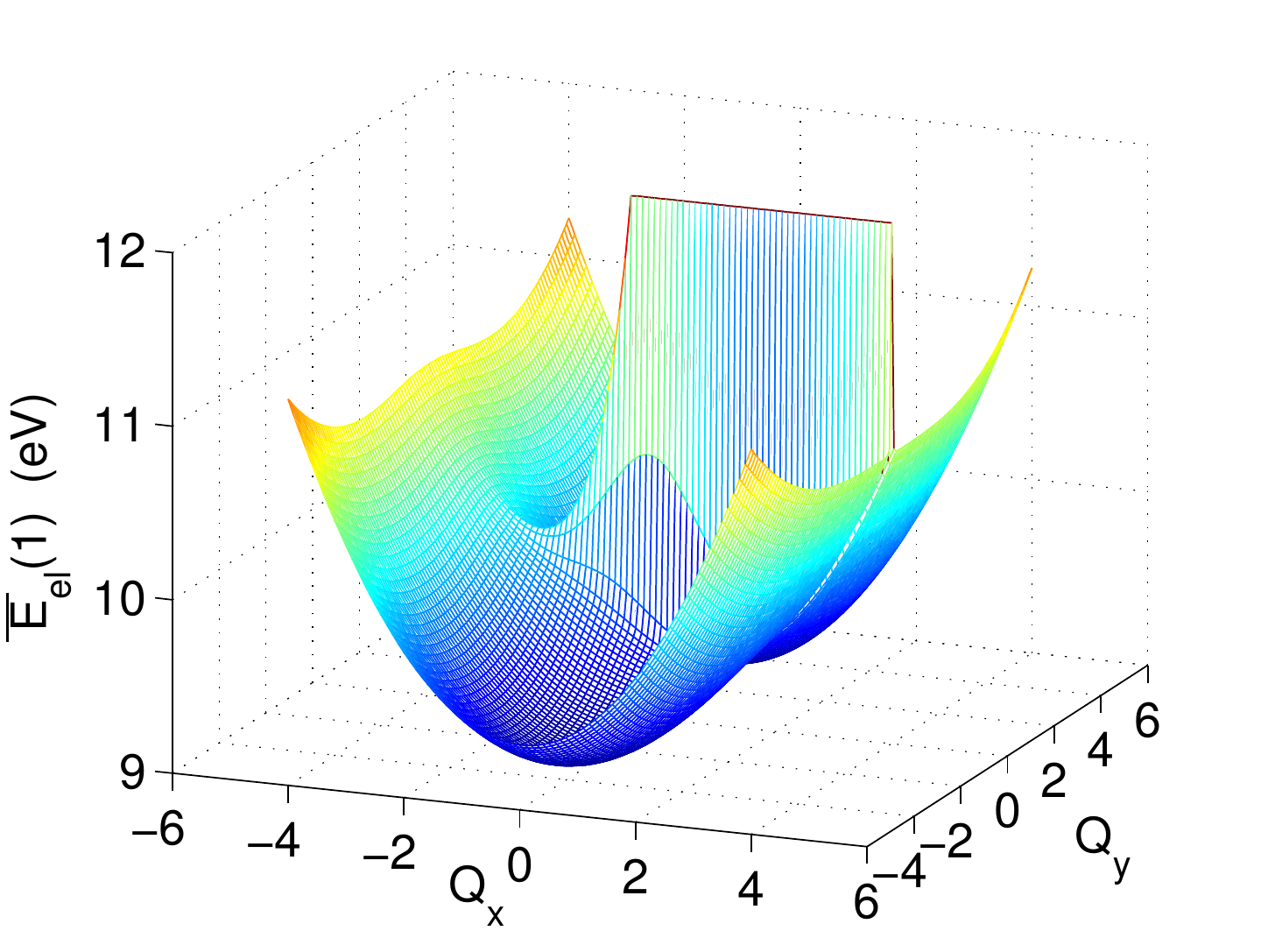}
 }
\subfigure[]{
   \includegraphics[width=6cm]{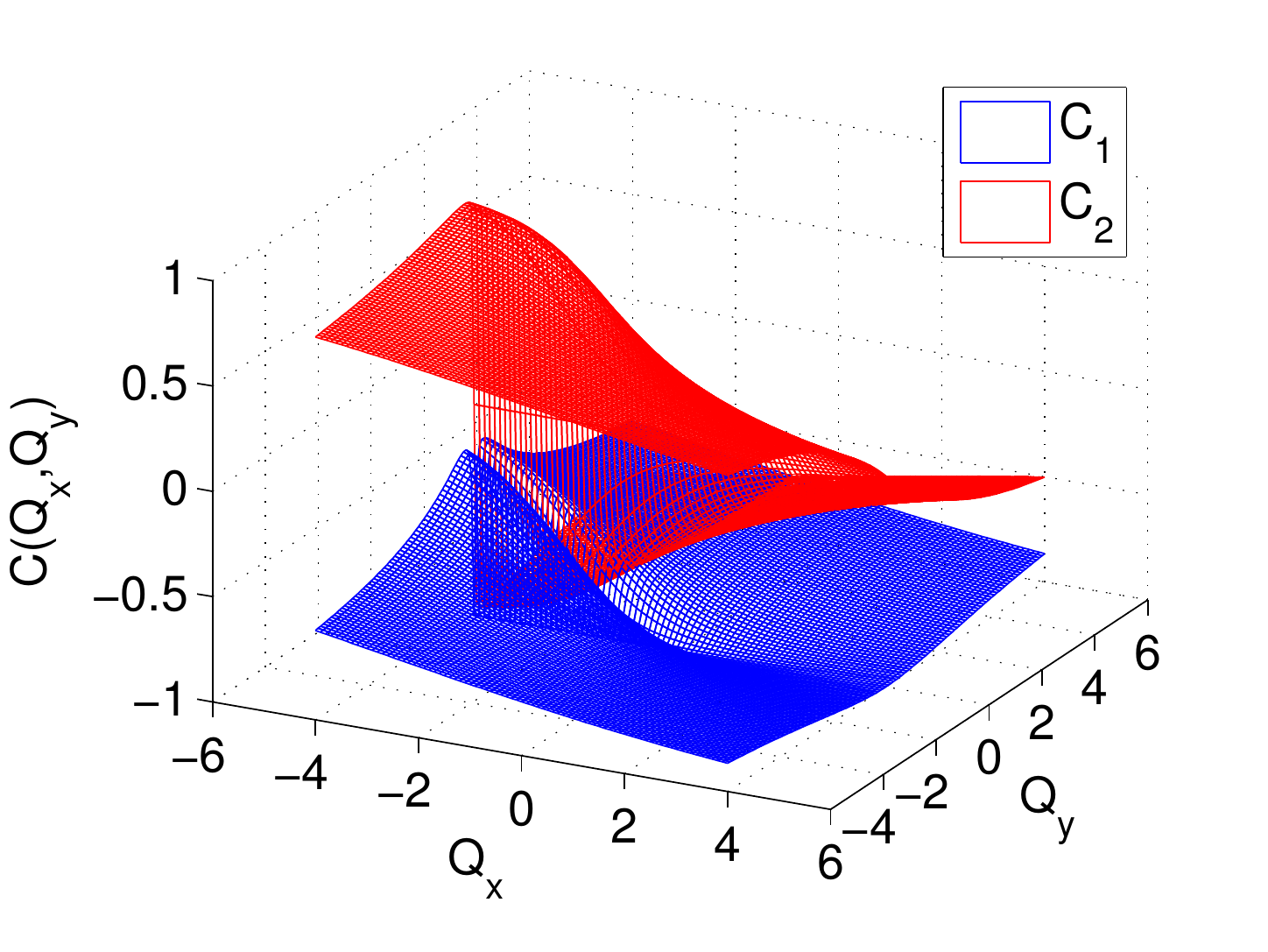}
 }
\subfigure[]{
   \includegraphics[width=6cm]{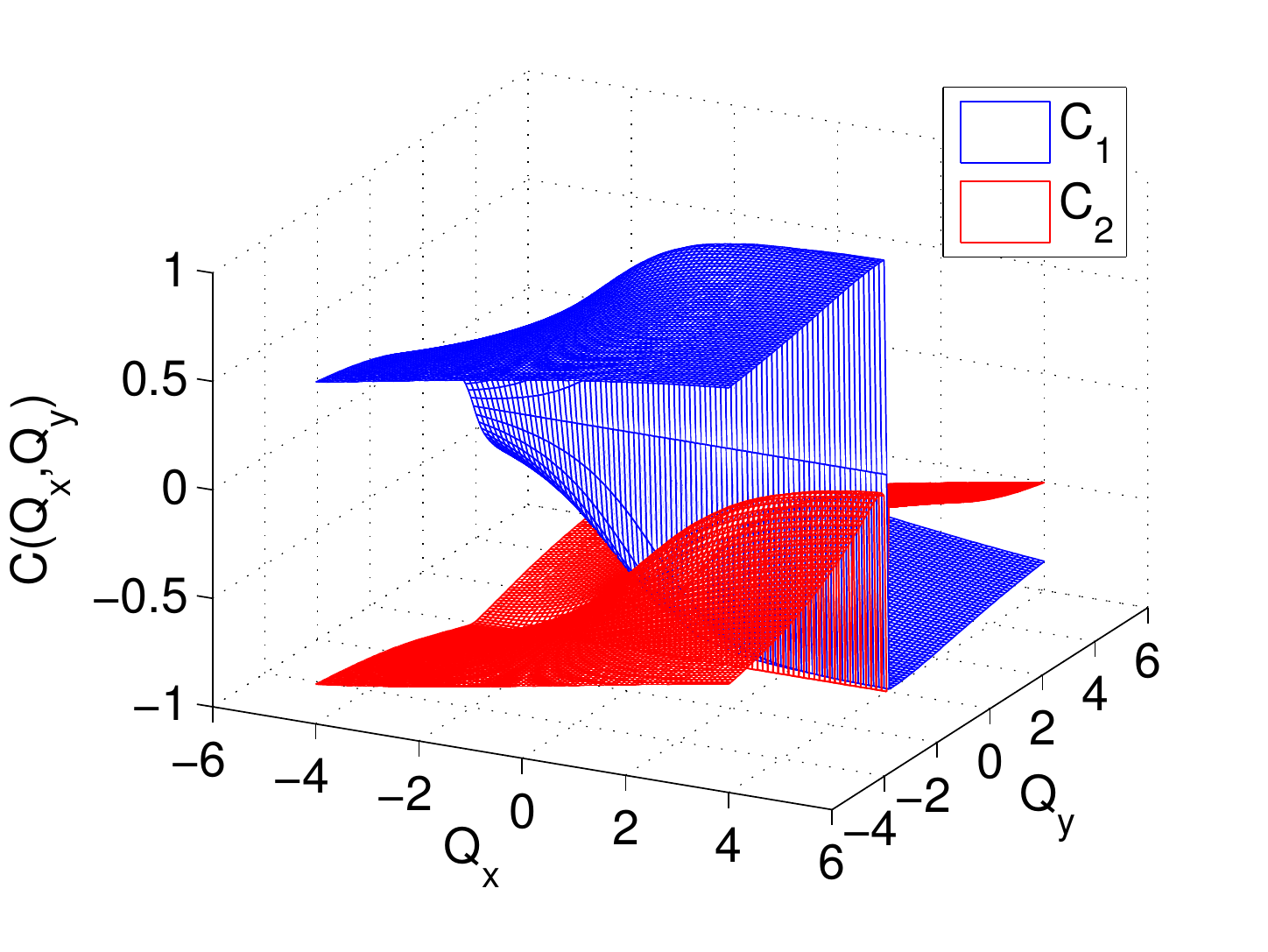}
 }
\caption{(Color online) Exact factorized potentials and coefficients of the first two 
vibrational states.  Panel (a) and (b): $\bar{E}_{\text{el}}(n)$ for $n=0$ and $n=1$, respectively. 
The lower adiabatic surface is the common struture of $\bar{E}_{\text{el}}(n)$,
and additionally there is a strong barrier in the middle, along $Q_y=0$.  In order to show the detail of the potential, the maximum value of the barrier is cut at 12 eV in both panels.
Panel (c) and (d): coefficients $C_1$ and $C_2$ for state $n=0$ and $n=1$, respectively.
In both panels, one of the coefficients changes sign along $Q_y=0$ which leads to the large 
barrier appearing in $\bar{E}_{\text{el}}(n)$.}
\label{fig:pot2dlow}
\end{figure}   

The exact potentials $\bar{E}_{\text{el}}(n)$ for $n=0,1$ and the corresponding coefficients
$C_1$ and $C_2$ are plotted in Fig.~\ref{fig:pot2dlow}.  Different from the previous example, 
the exact potentials now look like the \emph{adiabatic} potentials superimposed by large barriers.
Since our $\bar{\varphi}_n$ always follows the better electronic basis, our basic potential shape
in this example is, of course, the adiabatic potential.
These potential barriers, similar to our 1D example, are caused by the kinetic energy contribution 
of Eq.~\ref{eq:Hel2}, where one of the coefficients ($C_1$ and $C_2$) changes its sign, see e.g. panel 
(a) and (c).  One interesting phenomenon is
observed: this middle barrier appears alternatingly mainly at $Q_x < 0$ or at $Q_x > 0$
while $n$ increases.  
For the $\bar{E}_{\text{el}}^{(n)}$ of higher excited states, the lower
adiabatic potential remains the basic structure and so does the
middle barrier, even for $n > 10$ where the adiabatic approximation
is known to fail, cf. Fig.~\ref{fig:olad}.

\begin{figure}
\centering 
\hspace*{-0.5cm}
\includegraphics[width=5.5cm]{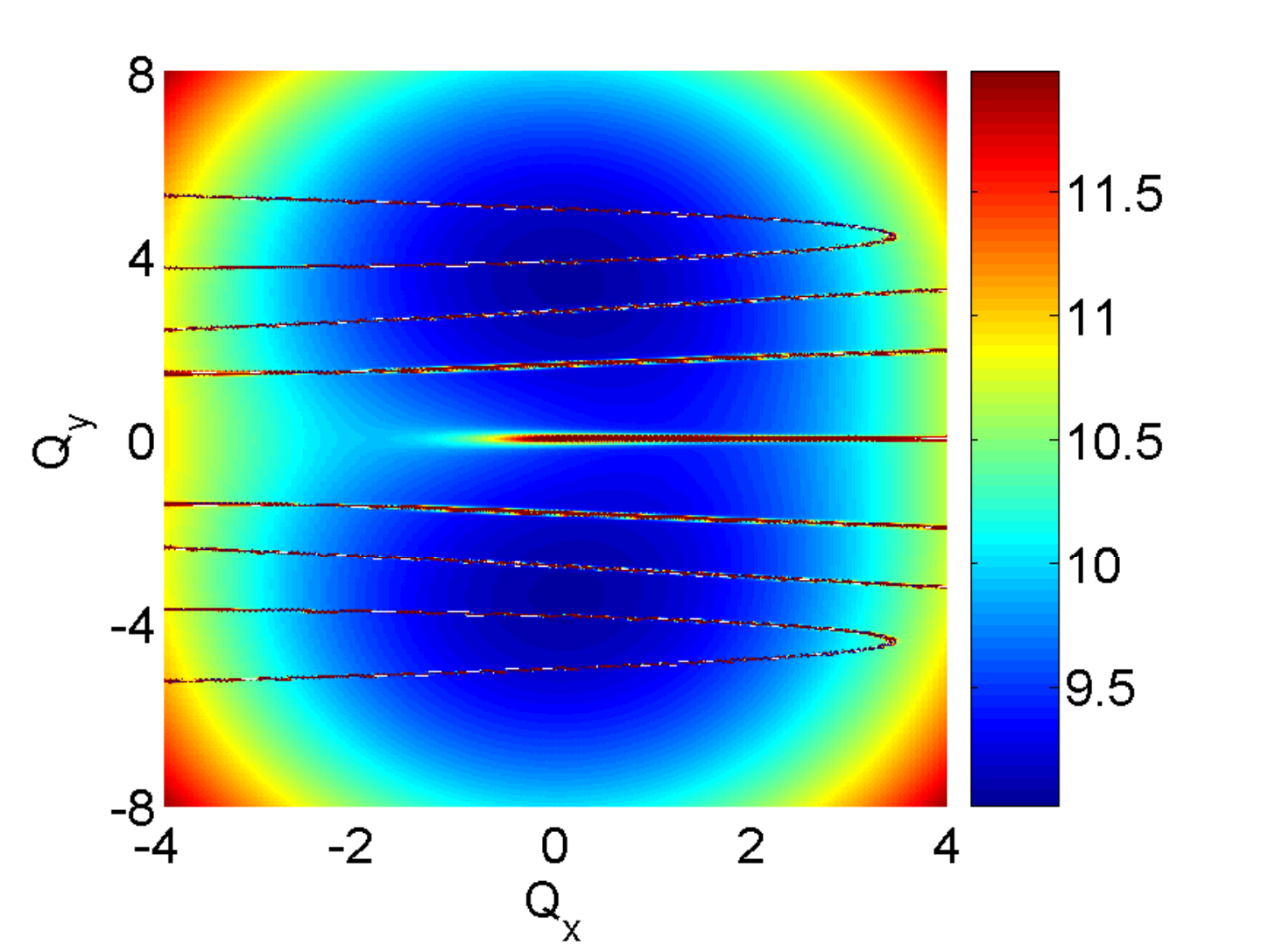}
\hspace*{-0.5cm}
\includegraphics[width=5.5cm]{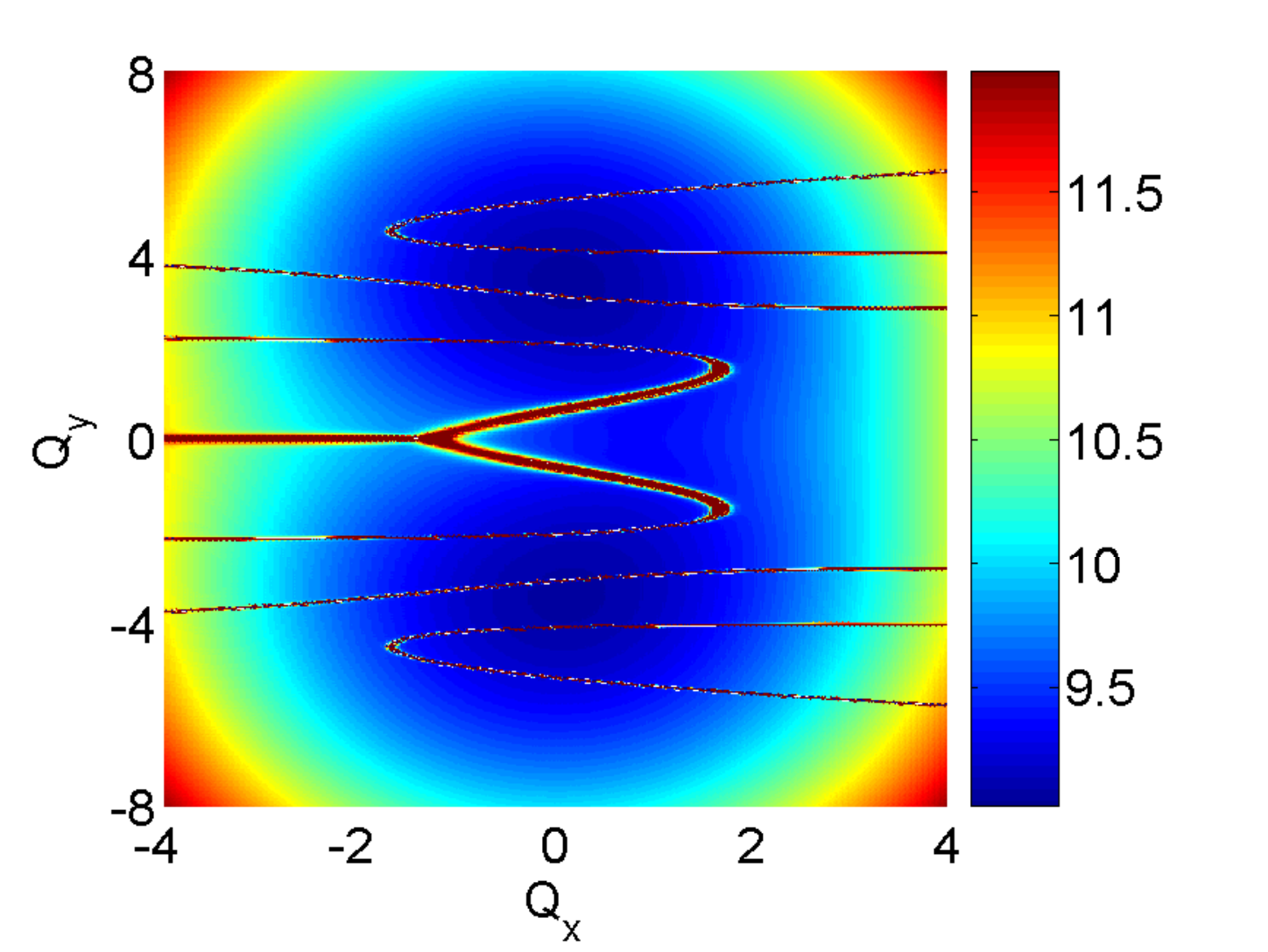}
\hspace*{-0.5cm}
\includegraphics[width=5.5cm]{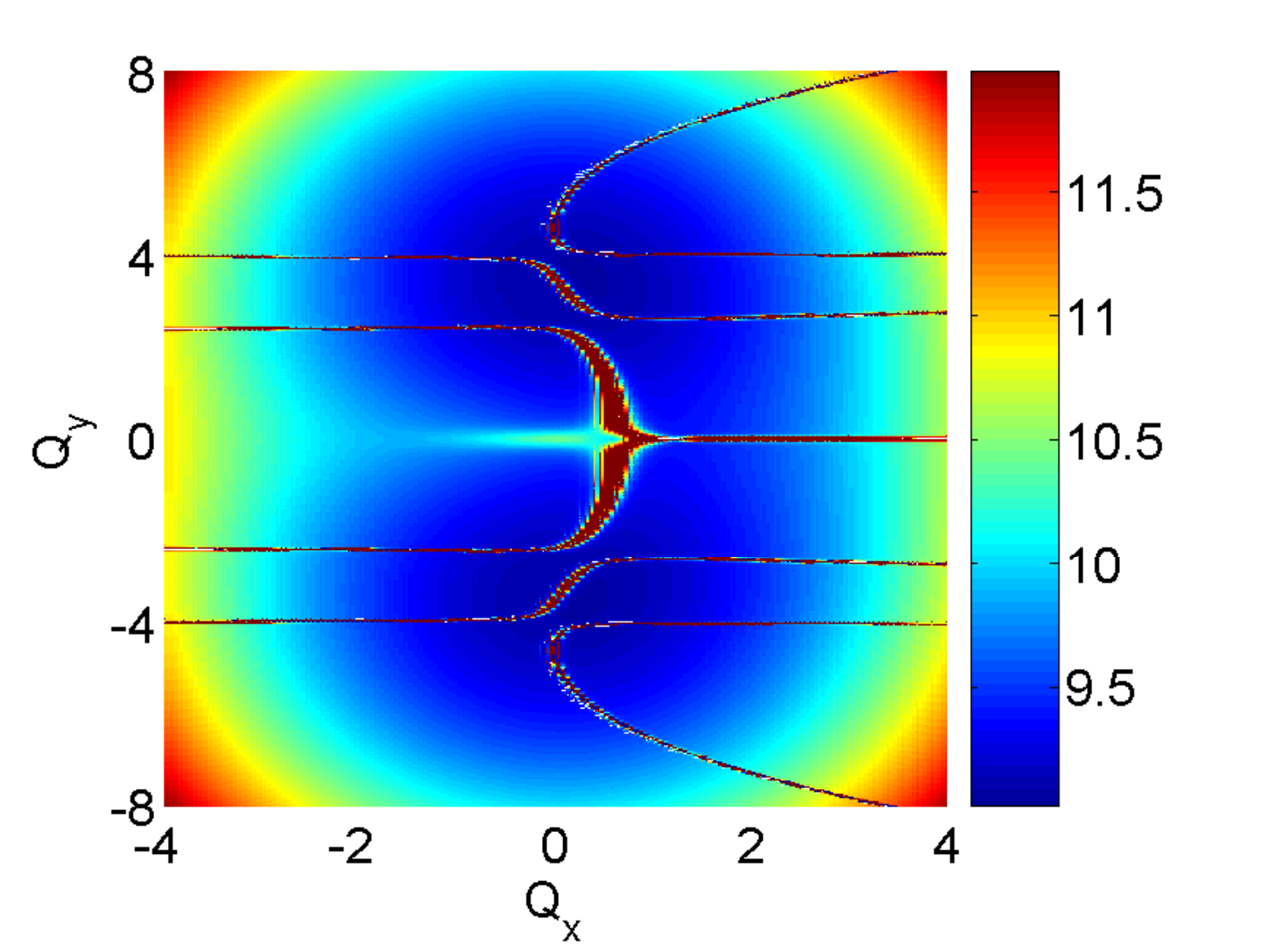}
\caption{(Color online) Potentials $\bar{E}_{\text{el}}(n)$. 
Shown in the panels from left to right are the potentials for $n=13$, $n=14$, and $n=15$,
respectively.  Again we cut the maximum of the poential barrier at 12 eV in order to 
present the details of the potential. The exact factorized potential is a superposition 
of the lower adiabatic surface and strong barriers.  Due to the double-well
symmetry, the middle barrier along $Q_y$ appears alternatingly at $Q_x < 0$ or $Q_x > 0$ 
when $n$ increases.}
\label{fig:pot2dhigh}
\end{figure}

\begin{figure}
\centering
\hspace*{-0.5cm} 
\includegraphics[width=4cm]{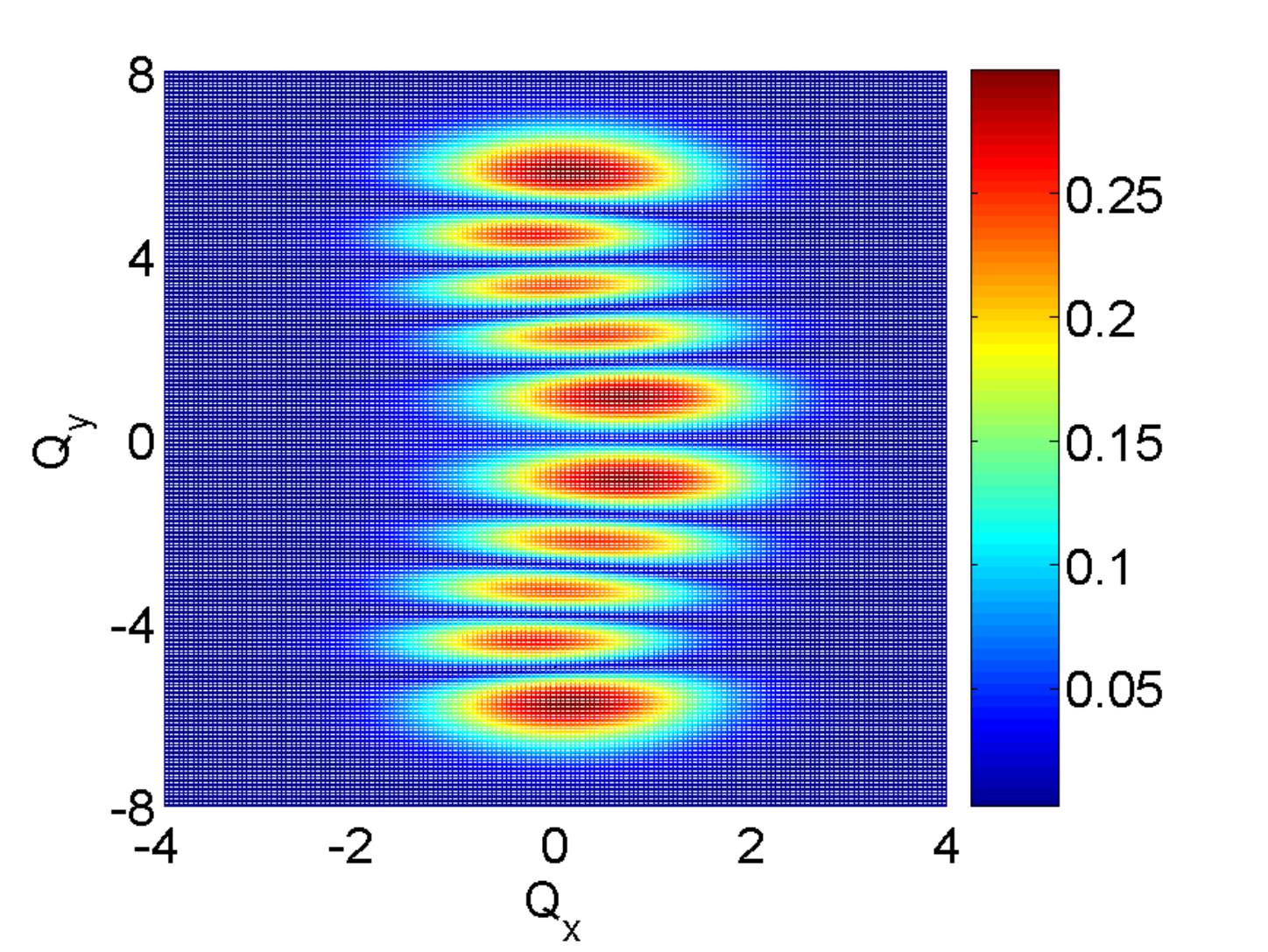} 
\hspace*{-0.5cm} 
\includegraphics[width=4cm]{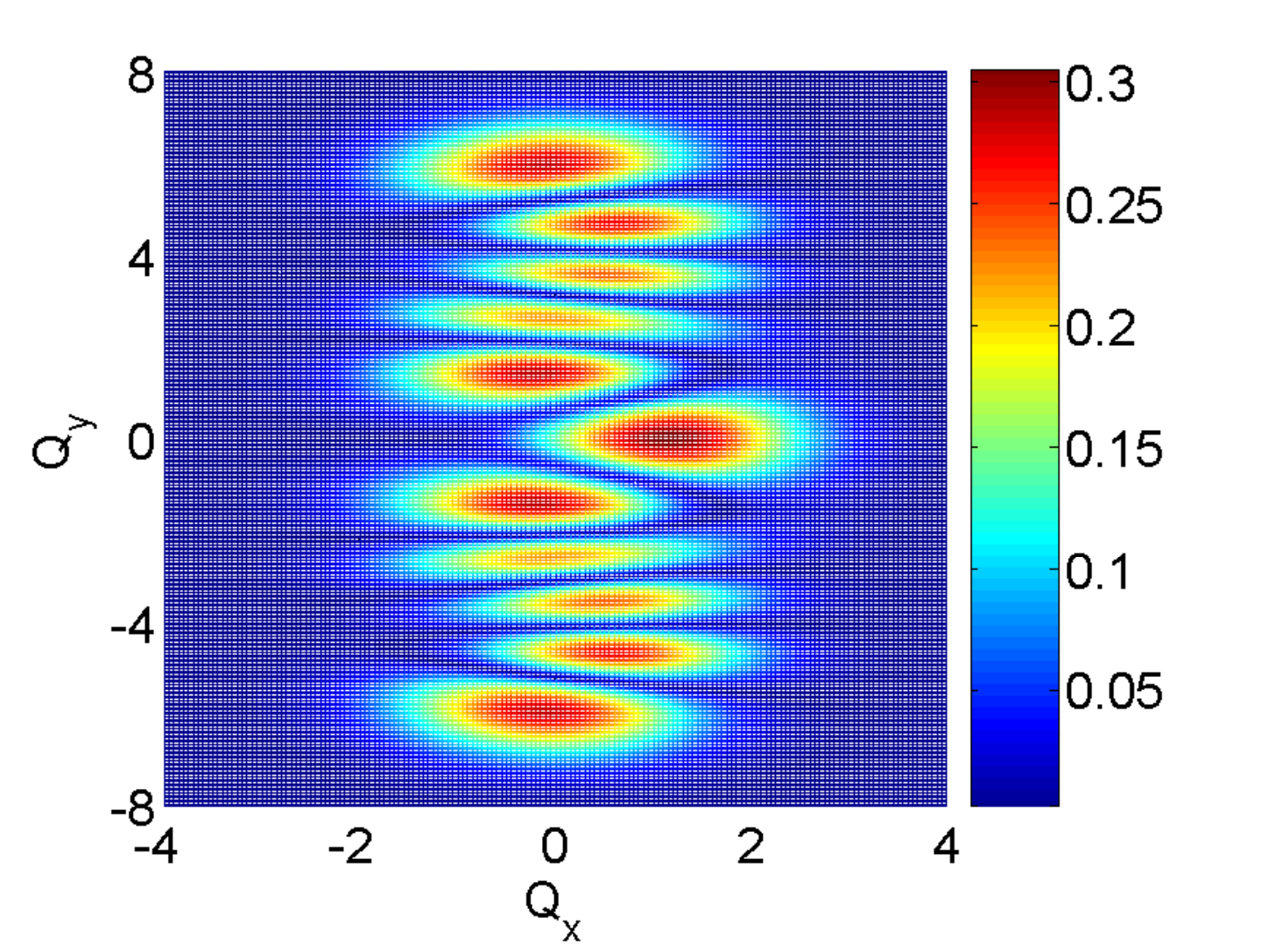}
\hspace*{-0.5cm}  
\includegraphics[width=4cm]{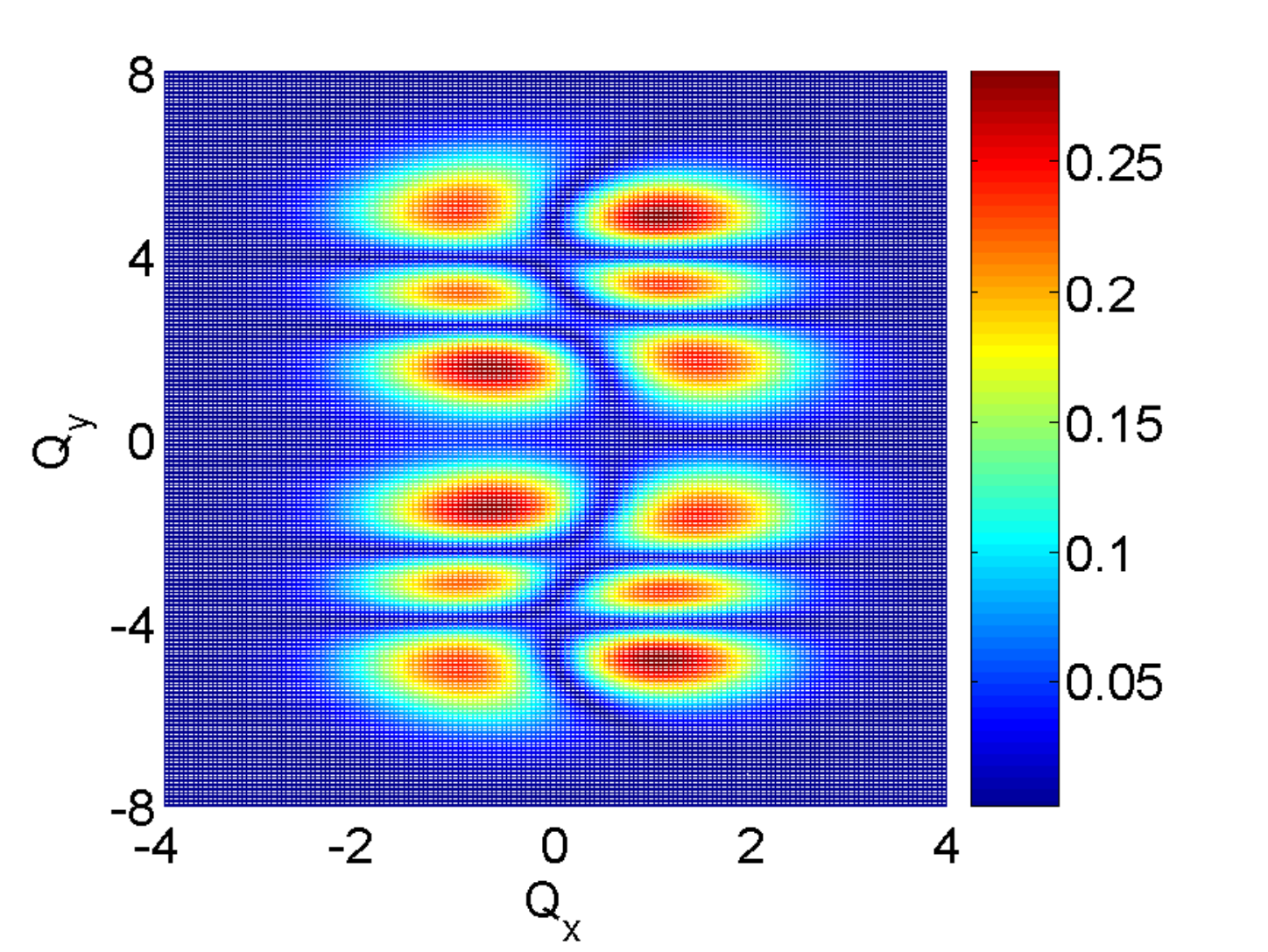}
\hspace*{-0.5cm}  
\includegraphics[width=4cm]{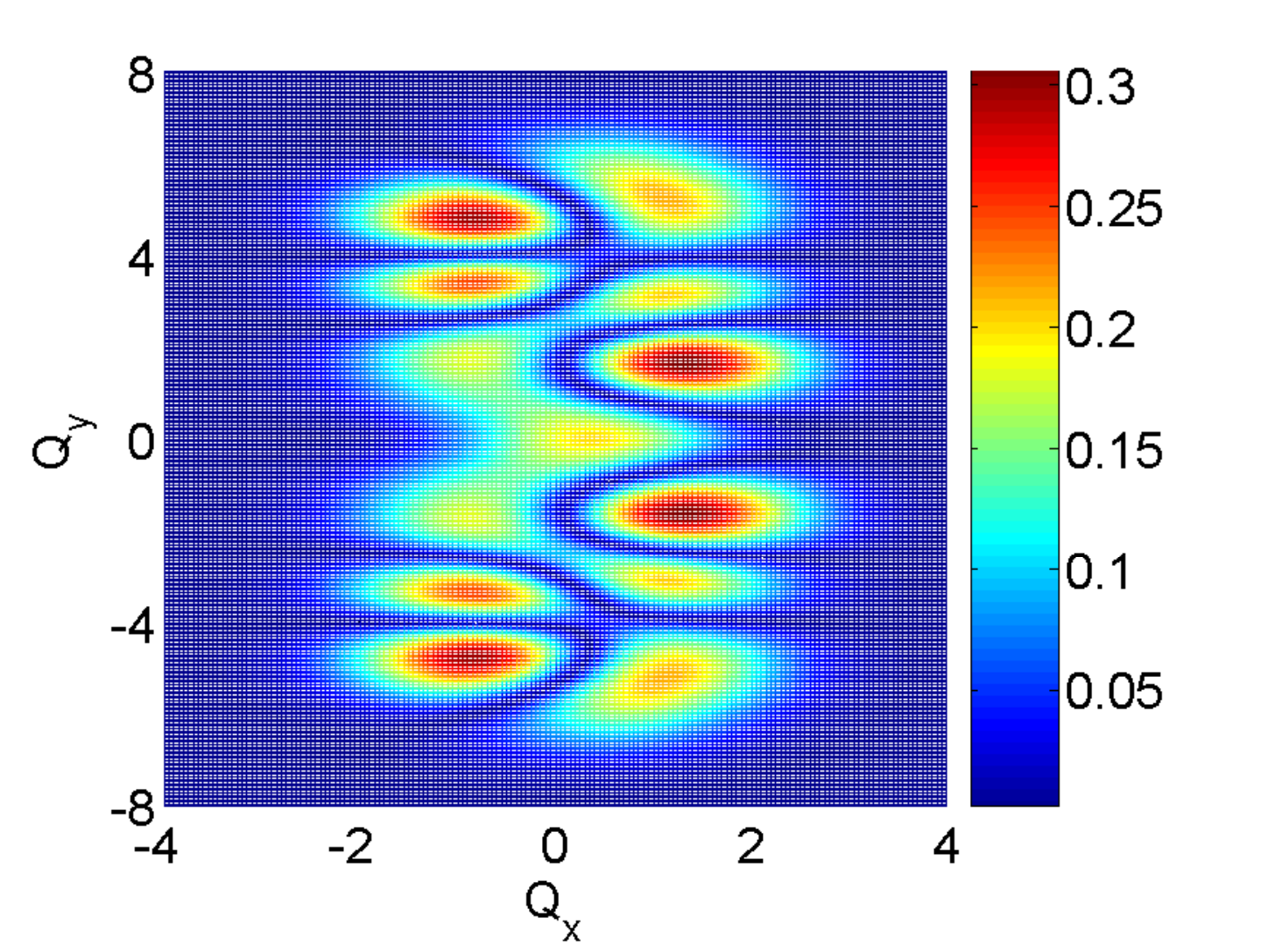}
\hspace*{-0.5cm}  
\includegraphics[width=4cm]{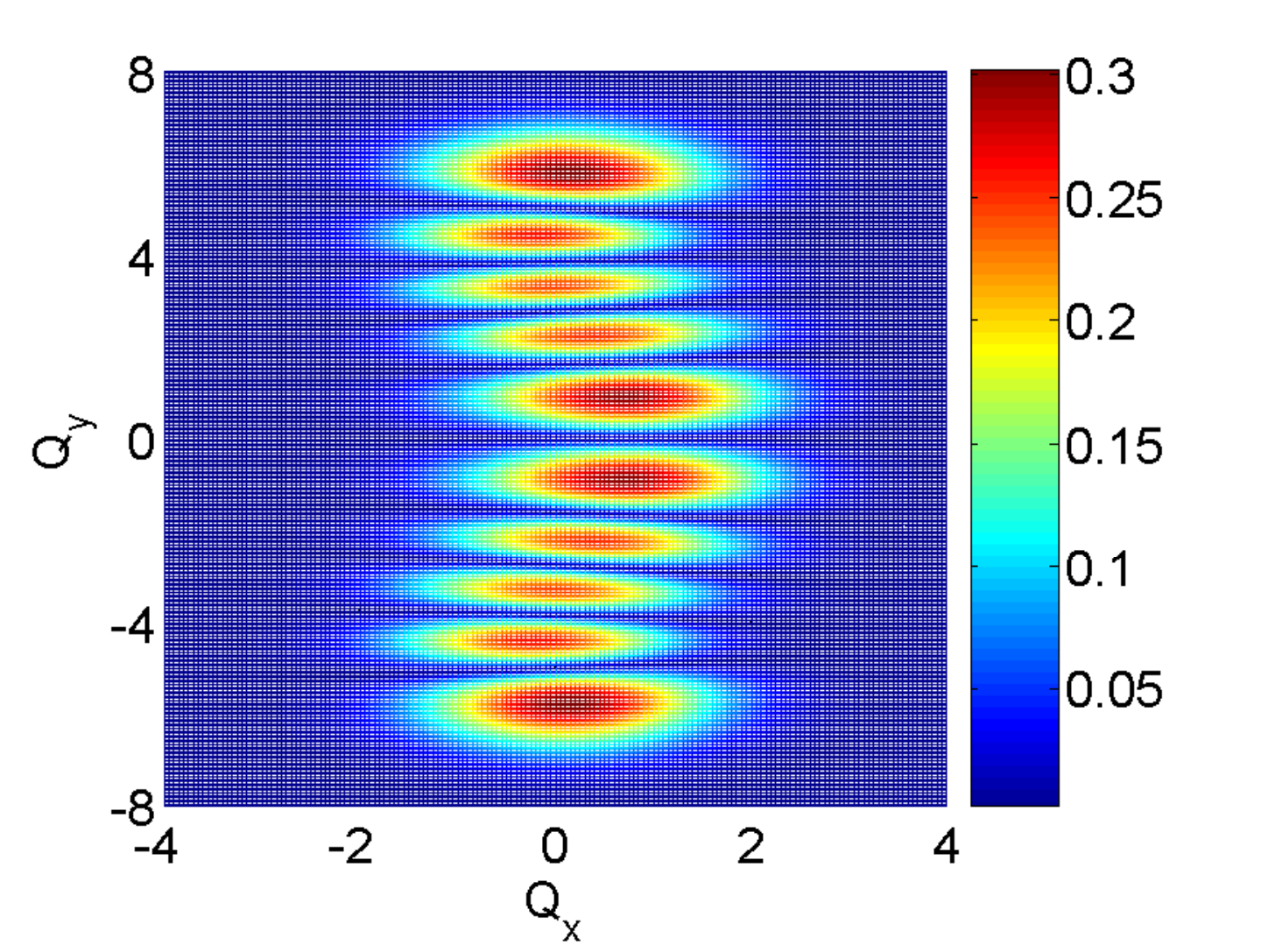} 
\hspace*{-0.5cm} 
\includegraphics[width=4cm]{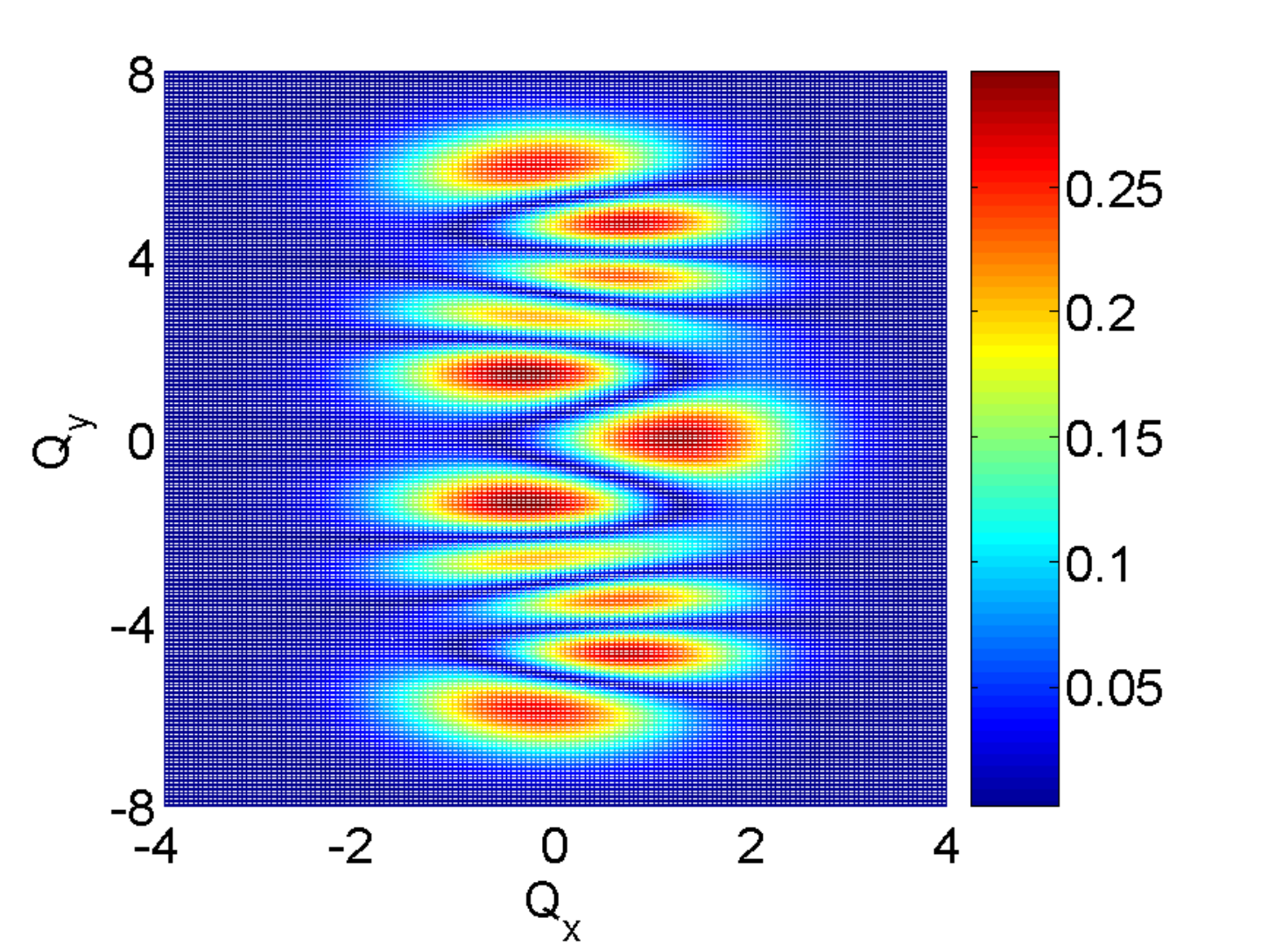} 
\hspace*{-0.5cm} 
\includegraphics[width=4cm]{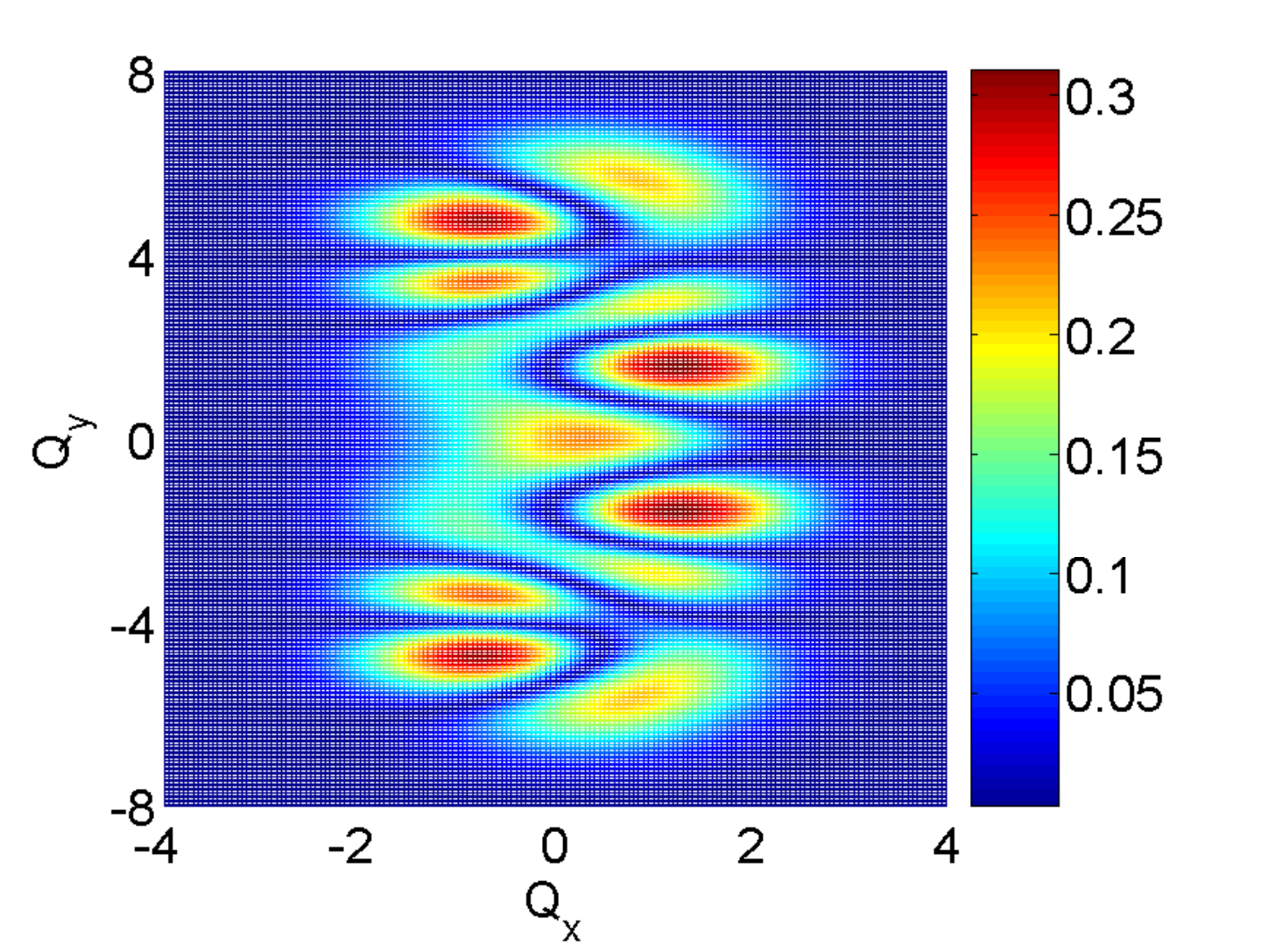}
\hspace*{-0.5cm}  
\includegraphics[width=4cm]{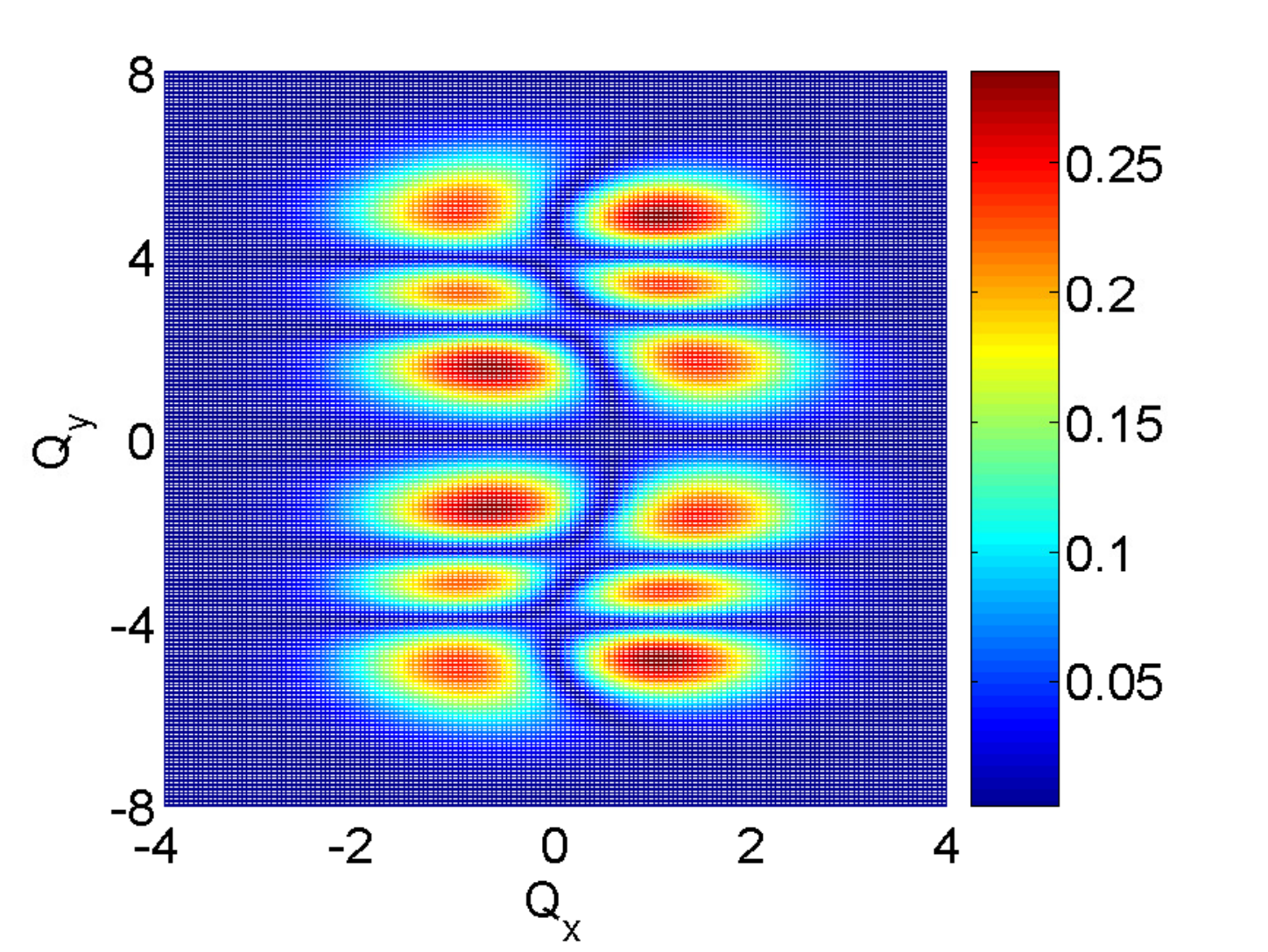}
\hspace*{-0.5cm}  
\caption{(Color online) Eigenfunctions of $\bar{H}_{\text{N}}^{(n)}$ and $H_{\text{ad}}$. 
Depicted in the top panels from left to right are $\bar{\chi}_n$ for $n=13-16$.  These are
nodeless nuclear eigenfunctions with multiple peaks, which are created by the potential barriers.  
The corresponding exact energy eigenvalues are 9.5875, 9.6347, 9.6413, and 9.6417 eV.  
Bottom panels from left to right:  $|\chi_{\text{ad}}^{(n)}|$ for $n=13-16$.
The corresponding adiabatic energy eigenvalues are  9.5860, 9.6334, 9.6375, and 9.6403 eV. 
Surprisingly they are almost identical with the corresponding $\bar{\chi}_n$;
only the order of last two ($n=15,16$) eigenfunctions should be exchanged.}
\label{fig:chibar2d}
\end{figure}

\begin{figure}
\centering 
\subfigure[]{ \includegraphics[width=7cm]{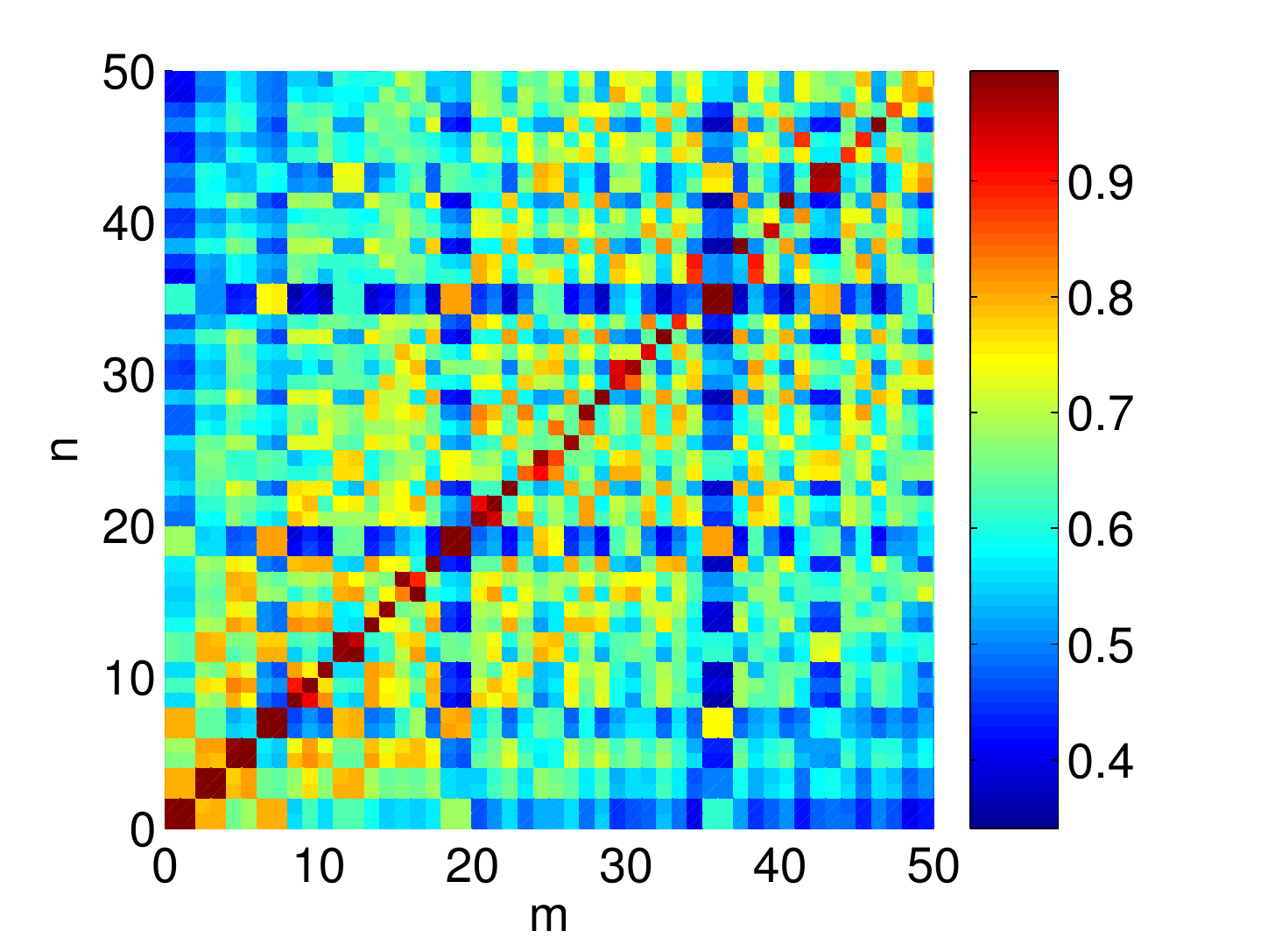} }
\subfigure[]{ \includegraphics[width=7cm]{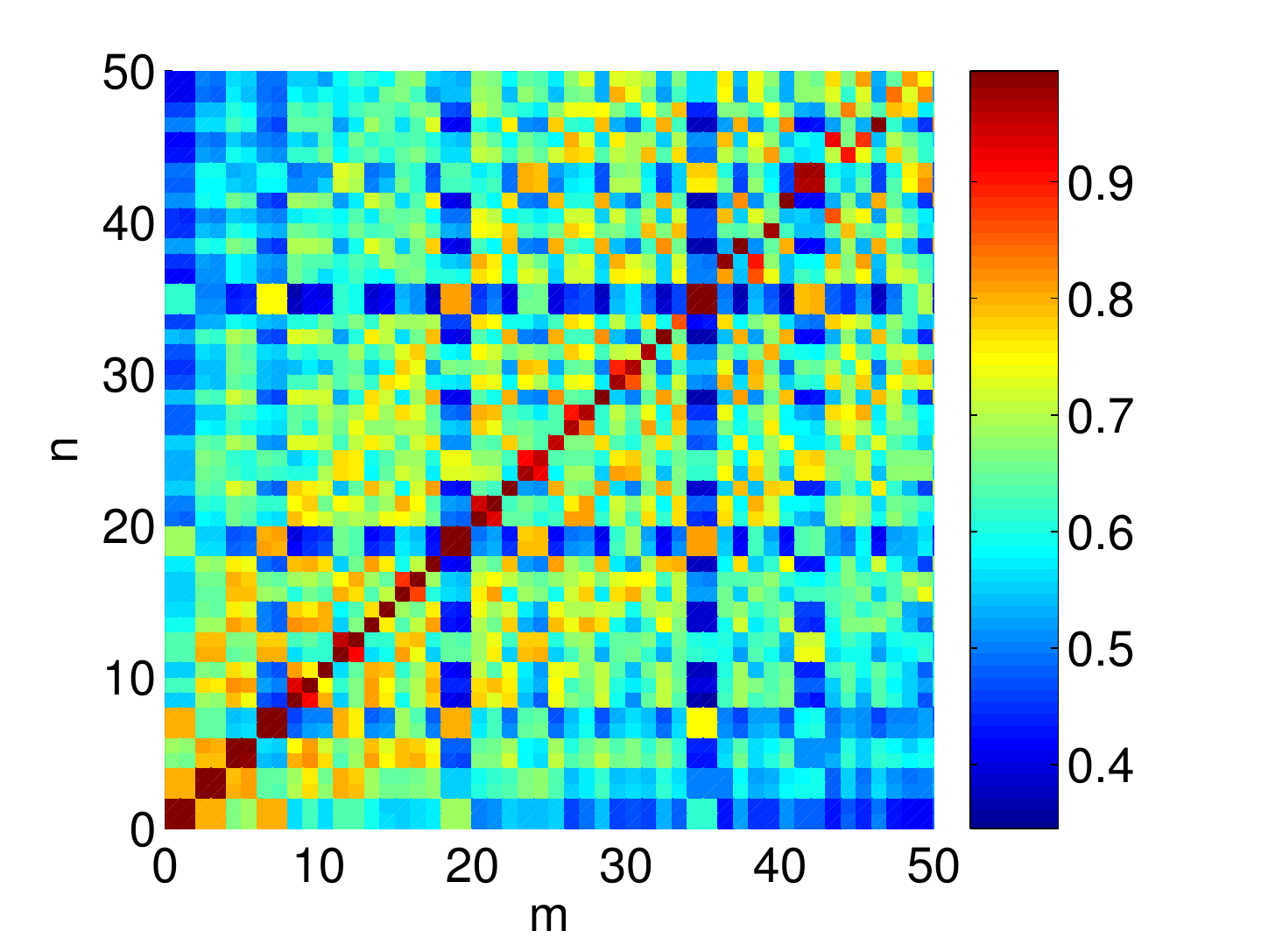} } 
\caption{(Color online) Panel (a): overlap matrix of the modulus of $\chi_{\text{ad}}$ ($m$) and $\bar{\chi}_n$ ($n$). 
Note that the energy now covers the range from 9.2381 eV to 10.0992 eV. Panel (b): overlap matrix of the modulus of $\chi_{\text{B.-H.}}$ ($m$) with $\bar{\chi}_n$ ($n$).  An overlap maximum almost continuously appears on the diagonal of the matrix, up to $n=50$. 
This demonstrates that the modulus of the adiabtic eigenfunction or of the Born-Huang adiabatic eigenfunction is an excellent approximation to the nodeless $\bar{\chi}_n$.}
\label{fig:absolad}
\end{figure}

For example, the exact potentials for $n$=13, 14, and 15 are depicted
in Fig.~\ref{fig:pot2dhigh}; all of them have the same double-well
surface with additional barriers superimposed.
These barriers, even though originating from the nodes in the vibronic eigenfunctions, 
somehow are very close to where the adiabatic eigenfunction has nodes. So, how good is the 
absolute value of an adiabatic eigenfunction, $|\chi_{\text{ad}}|$, as an approximation to 
$\bar{\chi}_n$?  In Fig.~\ref{fig:chibar2d}, $\bar{\chi}_n$ and $|\chi_{\text{ad}}|$
are plotted. In the upper panels the $\bar{\chi}_n$ for $n=13-16$ are given in ascending energy order, 
while in the lower panels the $|\chi_{\text{ad}}|$ are shown also in ascending energy order. 
One immediately sees that the modulus of an adiabatic eigenfunction
agrees well with $\bar{\chi}_n$, except that sometimes the
energy order of the adiabatic approximation has to be interchanged, cf. $n=15$ and $n=16$.
This feature certainly is remarkable, since the adiabatic eigenfunction for a long 
time was considered as meaningless when the adiabatic approximation fails. 
One might argue that this feature is not general and might break for higher energy 
eigenfunctions, e.g. for energy  around 10 eV.  Hence we show the
overlap matrix with elements $\langle |\chi_{\text{ad}}^{(m)}| | \bar{\chi}_n \rangle$
in Fig.~\ref{fig:absolad}.  Again, the maximum values of the overlap
matrix appear on the diagonal of the overlap matrix, and this feature seems
to continue up to n=50, whose energy is 10.0992 eV.
According to Fig.~\ref{fig:absolad},  the modulus of an adiabatic eigenfunction
is a good approximation to $\bar{\chi}_n$, even for the regime where the
adiabatic approximation fails.  In other words, $|\chi_{\text{ad}}|$
can be used as an initial guess for $\bar{\chi}_n$ in Eq.~\ref{eq:Hel}, when 
one solves the complete electron-nuclear coupled equations simultaneously.
Now, one should wonder which property actually goes wrong when 
the adiabatic approximation fails.  Why does the $|\chi_{\text{ad}}|$ follow so 
closely the exact $\bar{\chi}_n$ while simultaneously the overlap shown 
in Fig.~\ref{fig:olad} tells us that the adiabatic eigenfunctions do not well
represent the exact vibronic eigenfunctions?
Let us explain. 
An adiabatic eigenfunction, in contrast to $\bar{\chi}_n$,
has a phase/sign which depends on the nuclear degrees of freedom.
Additionally, the transformation matrix\cite{Horst84}  $\mathbf{S}$ 
which transforms the adiabatic basis into the diabatic basis also contains
a phase/sign. When the nuclear eigenfunction of the adiabatic basis is transformed
to the diabatic basis for comparison with the vibronic eigenfunctions,
cf. the overlap in Fig.~\ref{fig:olad}, the phase/sign can be different from
the vibronic one. In consequence, one gets an eigenfunction
which differs from the exact vibronic eigenfunction. 
Yet in our theory, this phase/sign should be included automatically
in the coefficients $C_1$ and $C_2$.  In our shortcut procedure, these
coefficients are obtained via employing the vibronic eigenfunctions, and 
hence the complete correlation between electrons and nuclei is treated correctly
for each vibronic state $n$.  One can, of course, take an approximate $C_1$
and $C_2$ where the electron-nuclei correlation is only partially treated.
A simple example is the adiabatic ground state eigenfunction.  If one
takes such an eigenfunction, the coefficents $C_1$ and $C_2$ are just
the adiabatic-to-diabatic transformation matrix elements $S_{11}$ and $S_{21}$, 
which are given analytically in Ref.~\cite{Horst84}.  
Inserting the coefficients $S_{11}$ and $S_{21}$ into Eq.~\ref{eq:Hel2},
the expectation value of the diabatic potential matrix yields the lower adiabatic
potential, while the expectation value of the kinetic energy operator matrix yields 
the usual diagonal correction term automatically! For derivation one can see Eq. 3.8 in Ref.~\cite{Horst84}.  
Therefore, if the adiabatic ground state eigenfunction is used for building approximate $\bar{E}_{\text{el}}^{(0)}$,
one obtains immediately the Born-Huang adiabatic approximation!
With this example, we conclude that the exact factorization wavefunction
ansatz \cite{Lenz13} is related to the adiabatic approximation but
allows a proper treatment of the full electron-nuclei correlation.



\section{Conclusion}

A total wavefunction ansatz given by a single product of an electronic and nuclear wavefunction is applied 
to one-mode and two-mode systems with non-adiabatic coupling. We employ diabatic electronic 
basis states with linear combination coefficients $C_1$ and $C_2$ to construct the exact factorized 
electronic wavefunction $\bar{\varphi}_n$.  These coefficients depend so strongly on the nuclear degrees 
of freedom that the exact potentials can be spiky.  
For the one-mode model, the exact potentials are given by the diabatic potential superimposed by spikes.
These spikes come from a rapid sign change of coefficients $C_1$ or/and $C_2$.
The exact potentials $\bar{E}_{\text{el}}^{(n)}$ of the two-mode butatriene example
are found to be the lower adiabatic surface superimposed by strong barriers.
The symmetry breaking effect due to the presence of the coupling mode is observed 
in $\bar{E}_{\text{el}}^{(n)}$ as well.  Due to the large barrier, $\bar{\chi}_n$ can be nodeless
but still have a multiple peak structure. We also find that the large barrier appears
close to the nodes of the adiabatic eigenfunctions.
More precisely, taking the modulus of the adiabatic nuclear eigenfunction yields an
extremely good approximation to  $\bar{\chi}_n$, even for those vibrational eigenstates
which cannot be described by the adiabatic approximation ($E_n>9.5$~eV) for the example
of butatriene. For further development, this discovery
indicates that the modulus of the adiabatic eigenfunctions is a good initial
guess for solving  the fully correlated time-independent Schr\"odinger equation of electron and
nuclei, especially for Eq.~\ref{eq:workingeq1} where an initial guess of $\bar{\chi}_n$
is required.  This modulus is, of course, not differentiable at the nodes of the adiabatic wavefunction,
but this problem can be solved via regularizing the wavefunction derivatives at the nodes. 
Additionally, our study shows the fundamental relation between the exact factorization
theory \cite{Lenz13} and the adiabatic and Born-Huang approximations. 
The exact factorization contains the complete electron-nuclei correlation in
each $\bar{\varphi}_n$, while the usual adiabatic approximation contains it only
partially, i.e. only the electron-nuclei attraction. If one uses the 
adiabatic approximation as an initial guess for the exact factorization method, 
the Born-Huang approximation is obtained.  
The latter still does not include the off-diagonal correction into the electronic state
and yields only good nuclear wavefunction amplitudes but not the correct phase/sign.
We also find that $\bar{H}_{\text{N}}^{(0)}$ is a better approximation 
than the adiabatic approximation, i.e. yielding better energies and 
eigenfunctions than the adiabatic ones. This provides hope that one can employ 
wavepacket propagation simulations with a single electronic surface when the energy range of interest
is suitable. Finally, we mention that the spatial imaging of individual
vibronic states is possible \cite{Huan11}.

To conclude, we show for the first time a systematic approach to apply
the exact factorization wavefunction ansatz to a conical intersection
system.  It allows us to investigate features like spiky potentials
and nodeless nuclear eigenfunctions.  Simultaneously it brings us a
deeper understanding of the adiabatic approximation, which yields
a good modulus of the nuclear wavefunction, but not necessarily the
correct phase/sign of the wavefunction, nor the correct ordering of the energy eigenvalues.

\section{Acknowledgement}

Y.C.C. thanks Prof. H. K\"oppel for many helpful discussions and the University of Heidelberg for financial support.
SK acknowledges the generous financial support of the Minerva foundation.

\end{document}